\documentclass{aa}
\usepackage{graphicx}
\usepackage{txfonts}
\usepackage{hyperref}
\usepackage{natbib}
\usepackage{lscape}

\newcommand{\kms}{km s$^{-1}$ }

\newcommand{\um}{$\mu$m }
\newcommand{\Tex}{$T\rm_{ex}$ }

\usepackage[normalem]{ulem}
\usepackage{color}

\begin{document}

\title{Kinematics and star formation toward W33: a central hub as a hub--filament system}
\titlerunning{CO OBSERVATION TOWARD W33}
\authorrunning {Liu et al.}
\author{Xiao-Lan Liu\inst{1,2}, Jin-Long Xu\inst{1,2}, Jun-Jie Wang\inst{1,2}, Nai-Ping Yu\inst{1,2}, Chuan-Peng Zhang\inst{1,2}, Nan Li\inst{1,2}, Guo-Yin Zhang\inst{1,2}}
\institute{National Astronomical Observatories, Chinese Academy of Sciences, Beijing 100101, China\\
\email{liuxiaolan@bao.ac.cn}
\and CAS Key Laboratory of FAST, National Astronomical Observatories, Chinese Academy of Sciences, Beijing 100101, China}

\date{Received  /
Accepted }

\abstract
{}
{We investigate the gas kinematics and physical properties toward the W33 complex and its surrounding filaments. We study clump formation and star formation in a hub--filament system.}
{We performed a large-scale mapping observation toward the W33 complex and its surroundings, covering an area of $1.3^\circ \times 1.0^\circ$ , in $^{12}$CO (1-0), $^{13}$CO (1-0), and C$^{18}$O (1-0) lines from the Purple Mountain Observatory (PMO). Infrared archival data were obtained from the Galactic Legacy Infrared Mid-Plane Survey Extraordinaire (GLIMPSE), the Multi-band Imaging Photometer Survey of the Galaxy (MIPSGAL), and the Herschel Infrared Galactic Plane Survey (Hi-GAL). We distinguished the dense clumps from the ATLASGAL survey. We used  the GLIMPSE I catalogue to extract young stellar objects.}
{We found a new hub--filament system ranging from 30 to 38.5 \kms located at the W33 complex. Three supercritical filaments are directly converging into the central hub W33. Velocity gradients are detected along the filaments and the accretion rates are in order of $\rm 10^{-3}\,M_\odot\, yr^{-1}$. The central hub W33 has a total mass of $\rm\sim 1.8\times10^5\,M_\odot$, accounting for $\sim 60\%$ of the mass of the hub--filament system. This indicates that the central hub is the mass reservoir of the hub-filament system. Furthermore, 49 ATLASGAL clumps are associated with the hub--filament system. We find $57\%$ of the clumps to be situated in the central hub W33 and clustered at the intersections between the filaments and the W33 complex. Moreover, the  distribution of Class I young stellar objects (YSOs)  forms a structure resembling the hub--filament system and peaks at where the clumps group; it seems to suggest that the mechanisms of clump formation and star formation in this region are correlated.  }
{Gas flows along the filaments are likely to feed the materials into the intersections and lead to the clustering and formation of the clumps in the hub--filament system W33. The star formation in the intersections between the filaments and the W33 complex might be triggered by the motion of gas converging into the intersections.}
\keywords{stars: formation - stars: kinematics and dynamics - stars: massive - (ISM:) HII regions - ISM: lines and bands - ISM: Molecules}
\authorrunning{Liu et al.}

\maketitle

\section{Introduction}
High-mass stars ($> 8$ M$_\odot$) play a vital role in the evolution of interstellar medium (ISM) and their host galaxies. They can enrich the chemical components in the ISM by throwing heavy metal elements into the ISM and therefore promote the chemical evolution of galaxies. However, the formation mechanism of high-mass stars remains unknown. Recently, filaments have found to be prevalent as the main sites of star formation \citep[e.g.][]{Myers09,Andre10,Andre14,Wang15}. Particularly in the hub--filament systems, active high-mass star formation is frequently reported in hubs, which host a number of clumps in different phases \citep[e.g.][]{Schneider12,Peretto13,Hennemann14,Friesen16,Yuan18}. In addition, theoretical studies suggest that filaments in the hub--filament systems may act as tributaries, converging mass into the protocluster clumps in the central hub and therefore resulting in vigorous star formation there \citep{Myers09,Liu12,Gomez14}. Therefore, identifying and investigating the hub--filament systems can help us to understand the formation of clumps and massive stars.

The W33 complex is a massive star-forming region located at  $l\sim 12^\circ.8$ and $b\sim -0.2^\circ$ in the Galactic plane. The parallactic distance of the W33 complex is $\rm 2.4^{+0.17}_{-0.15} \,kpc$ \citep{Immer13}. The W33 complex contains six massive dust clumps (W33 Main, W33A, W33B, W33 Main1, W33A1, and W33B1), which were identified by \citet{Immer14} using the APEX Telescope Large Area Survey of the Galaxy (ATLASGAL) \citep{Schuller09}. The calculations toward these clumps suggest a total mass of $\rm\sim (0.8-8)\times 10^5 \,M_\odot$ and an integrated IR luminosity of $\rm \sim 8\times 10^5\, L_\odot$ \citep{Immer14, Kohno18}. Moreover, molecular line observations toward the W33 complex detect a complex velocity field \citep[e.g.][]{Gardner72,Goldsmith83,Immer13,Kohno18}. As for W33A and W33 Main, emission and absorption peaks were observed at a radial velocity of $\sim 35$ \kms, while the spectra toward W33B were seen to peak at $\sim$ 60 \kms \citep{Gardner72,Goldsmith83}. \citet{Immer13} proved that these two velocity components are located at the same parallaxial distance using the water maser emission. Apart from the 35 \kms and 58 \kms velocity components associated with the W33 complex, \citet{Kohno18} detected and identified another velocity component at 45 \kms using the CO data from the NANTEN2 and Nobeyama 45 m telescopes. Furthermore, the 21 cm absorption line data suggest that this velocity component is likely partly associated with the W33 complex \citep{Kohno18}.  \citet{Kohno18} propose that the cloud--cloud collision scenario between the 35 \kms and 58 \kms clouds can explain the observed properties.

Visualisation of the W33 complex on a large scale with the three-colour composite of 8, 24, and 70 $\mu$m in Fig. 1 suggests that the W33 complex is similar to a central hub surrounded by a set of infrared-dark filaments. In order to highlight the kinematics in the W33 complex and the effects that the surroundings have on the W33 complex, we performed a large-scale CO molecular line observation towards the W33 complex. In addition, the archival infrared data are adopted to probe star formation activities. In Section 2, we describe the data sets, and in section 3 the results. We present our analysis and discussions in Section 4. Finally, we summarise our main results in Section 5.

\section{Observations and data reduction}
\subsection{Purple mountain data}
The mapping observations were made towards the W33 complex and its adjacent regions in $^{12}$CO (1-0), $^{13}$CO (1-0), and C$^{18}$O (1-0) lines (the rest frequencies of 115.10, 110.20, and 109.78 GHz) using the PMO 13.7 m radio telescope at De Ling Ha in western China at an altitude of 3200 m during October 2017. The total mapping extent was approximately $80'\times60'$ for all the lines. The half-power beam width (HPBW) at 115 GHz is $\sim 53''$ \footnote{http://english.dlh.pmo.cas.cn/fs/}. The new nine-beam array receiver system in single-sideband mode (SSB) was used as a front end. Fast Fourier transform spectrometers were used as a back end with a total bandwidth of 1 GHz and 16384 channels. $^{12}$CO (1-0) was observed at upper sideband with a system noise temperature ($T\rm_{sys}$) in range of $150-300$ K, while $^{13}$CO (1-0) and C$^{18}$O (1-0) were observed simultaneously at lower sideband. The velocity resolution for $^{12}$CO (1-0) is $\sim$ 0.16 \kms and for $^{13}$CO (1-0)/C$^{18}$O (1-0) is $\sim$ 0.17 \kms. The pointing accuracy of the telescope was better than $4''$. The phase centre of the observations was $(l, b)$ $\sim$ (13$^\circ$.222, -0$^\circ$.219) with the position-switch on-the-fly (OTF) mode. The off-position was $(l, b)$ $\sim$ (12$^\circ$.375, -2$^\circ$.124), devoid of CO emission. The standard chopper wheel calibration technique was used to measure the antenna temperature T$\rm_A^*$ corrected for the atmospheric absorption. The final data were recorded in a brightness temperature scale of T$\rm_{mb}$ (K). The data were reduced and regridded using the software GILDAS \citep{Pety05}. The pixel sizes of the CO FITS cubes were $30''\times30''$ . The $^{12}$CO (1-0) data are merely used to estimate the excitation temperatures (see Appendix A), because the velocity components in the $^{12}$CO (1-0) spectra cannot be disentangled.

\subsection{Archival data sets}
\subsubsection{Spitzer space telescope}
The Galactic Legacy Infrared Mid-Plane Survey Extraordinaire (GLIMPSE) observed the Galactic plane ($10^\circ<|l|<65^\circ$, $|b|<1^\circ$) with the Infrared Array Camera (IRAC) instrument that can obtain simultaneous broadband images at 3.6, 4.5, 5.8, and 8 \um \citep{Fazio04}. We downloaded the mid-infrared (MIR) image at 8 \um of this region from the \emph{Spitzer} GLIMPSE \citep{Benjamin03}. The archived angular resolutions of IRAC 3.6, 4.5, 5.8, and 8.0 \um are $1.7''$, $1.7''$, $1.9''$, and $2.0''$, respectively. In addition, we also extracted the MIPSGAL 24 \um image covering the same region with a resolution of $\sim 6''$ \citep{Carey09}. The point-source catalogue released from the GLIMPSE was used in the following analysis.
\subsubsection{ATLASGAL sources}
\citet{Csengeri14} extracted $\sim$ 10861 compact submillimetre sources with fluxes above 5$\sigma$ from the maps of the ATLASGAL Survey  \citep{Schuller09}. The ATLASGAL Survey was conducted by the Large APEX Bolometer Camera (LABOCA) at the 12m diameter APEX telescope at 870 \um \citep{Gusten06a,Gusten06b}, dominated by the cold dust emission from dense clumps. It was the first systematic survey of the inner Galactic plane. It has a beam size of $\sim \, 19''.2$ and in total covers 420 square degrees of the Galactic plane in the Galactic longitude region of $-80^\circ<l<60^\circ$.

\section{Results}
\subsection{Velocity components toward the W33 complex}
To fit the spectrum of each pixel in the $^{13}$CO (1-0) and the C$^{18}$O (1-0) fits cubes, we use Behind the Spectrum (BST)\footnote{https://github.com/SeamusClarke/BTS} \citep{Clarke18}, which is a fully automated routine. A detailed description of BST is given by \citet{Clarke18}. For each component, the BST returns the amplitude (i.e. peak intensity), centroid velocity ($\upsilon\rm_c$), velocity dispersion ($\sigma\rm_c$), and the reduced $\chi^2\rm_c$ values of each spectrum. For accuracy, we check all the fits, and manually remake the Gaussian fits to each spectrum. If it is fitted not well, we will exclude it. And the returned amplitudes have to be over 1 K ($5\sigma$) for C$^{18}$O (1-0) and 3 K ($6\sigma$) for $^{13}$CO (1-0). The velocity dispersions $\sigma\rm_c$ have to be more than 0.51 \kms (3 channels). Moreover any two fits from the same spectrum have a $\upsilon\rm_c$ separation greater than 0.51 \kms. Figure 2 shows the returned $\upsilon\rm_c$ histograms of the C$^{18}$O (1-0) and $^{13}$CO (1-0) fits, respectively. From the $\upsilon\rm_c$ distributions, we identify six velocity components along the line of sight toward the W33 complex. These are in the ranges of 2-15 \kms, 15-21 \kms, 21-28 \kms, 30-44 \kms, 44-48 \kms, and 48-60 \kms, marked with the black dashed boxes in Fig. 2. In addition, we estimate the excitation temperatures and the optical depths for the fits of two lines via equations A.1-A.2. The H$_2$ column densities for the C$^{18}$O fits are also calculated using equation A.3. The distribution diagrams of these parameters are shown in Figs. A.1-A.5. From the figures, we can see that the $^{13}$CO lines are really optically thick in the dense regions with $\rm N(H_2)\gtrsim 10^{23}\,cm^{-2}$. Here, we use the optically thin C$^{18}$O (1-0) line to determine which velocity component is associated with the  W33 complex.

Based on the above six velocity ranges, we make the C$^{18}$O (1-0) integrated intensity maps overlaid on the \emph{Spitzer} 8 \um emission, as shown in Fig. 3. The C$^{18}$O (1-0) emission  reveals completely different structures for the six velocity components. Also, from the morphologies of the C$^{18}$O (1-0) emission, we find that the velocity components of 30-44 \kms and 48-60 \kms seem associated with the W33 complex and W33B, respectively. This result is consistent with those of \citet{Immer13} and \citet{Kohno18}. On a larger scale, our CO velocity components reveal a number of other structures besides the W33 complex. In particular, the 30-44 \kms velocity component in Fig. 3d is closely associated with a set of infrared dark filaments, referred to here as f1-f5, which seem to surround the W33 complex. Below, we further use the position-position-velocity (PPV) cubes to discern the velocity structures associated with the W33 complex in the 30-44 \kms range.

\subsection{The hub--filament system}
\subsubsection{A new hub--filament system identified toward the W33 complex}
To highlight the velocity structures in 30-44 \kms, we make the C$^{18}$O (1-0) PPV space using the BST obtained fits cube, as shown in Fig. 4. According to the values of the centroid velocities $\upsilon\rm_c$, we mark the colour for each point. Meanwhile, the 3D projections of all the points are plotted on each of the axes. On each projection plane in Fig. 4, we identified three small velocity ranges, namely 30-38.5 \kms, 38.5-42 \kms, and 42-44 \kms. Figure 5 shows the centroid velocity distribution diagrams for the above three velocity ranges. From Fig. 5, we find that the C$^{18}$O (1-0) emission presents different structures in the three ranges. The structure of the 38.5-42 \kms velocity component in Fig. 5b mainly presents an east-to-west large-scale filament containing filaments f4 and f5. From Fig. 5c, we see that the velocity component of 42-44 \kms is mainly coincident with some isolated structures. While the velocity component of 30-38.5 \kms is closely associated with the W33 complex and filaments f1-f3; below we focus on this component. In Fig. 5a, an ellipse represents the W33 complex. To avoid confusion, we rename the filaments adjacent to the W33 complex as hf1, hf2, and hf3. The three filaments traced by the black lines gather in the direction of the W33 complex. Similar to the structure of NGC 2264  \citep{Kumar2020}, the 30-38.5 \kms velocity component appears to show a hub--filament system, whose central hub is the W33 complex. 

\subsubsection{Physical properties of the hub--filament system}
Furthermore, we estimate the physical parameters of the hub--filament system, which are all listed in Table 1. From column 12 of Table 1,  we see that the central hub W33 has an equivalent radius of 6.0 pc. The filament hf2 is the longest with a length up to 15.7 pc, while hf1 is the shortest with a length of 7.5 pc. The widths of the filaments hf1, hf2, and hf3 are 4.6 pc, 3.8 pc, and 3.0 pc, respectively. Considering the sensitivity limits of C$^{18}$O (1-0) ($\rm Amplitude > 5\sigma$) as well as the inclinations, the measured lengths and widths of the filaments should be a lower limit. Column 4 of Table 1 presents the velocity ranges  spanned by these structures. The minimum velocity interval is $\sim 5$ \kms in the filament hf3, further demonstrating the complicated velocity structures in the whole hub--filament system. In addition, column 7 in Table 1 shows the minimum ratios of the non-thermal velocity dispersions to the sound speeds ($ \sigma_{\rm NT}/c\rm_s $) in these structures. We can see that all the ratios $ \sigma_{\rm NT}/c\rm_s $ are greater than 1, suggesting that the whole hub--filament system is supersonic, probably dominated by the turbulence \citep{Liu18}. The hub--filament system has a mean excitation temperature of $\sim17.5$ K, less than that of the W33 complex ($\sim20.7$ K), indicating that the central hub is warmer than its surrounding environment. Also, from columns 8-9 of Table 1, both the central hub W33 and the filaments hf1, hf2, and hf3 have a mean H$_2$ column density in magnitude of 10$^{22}$ cm$^{-2}$ and a mean H$_2$ number density on the order of 10$^4$ cm$^{-3}$. In addition, the W33 complex has a total mass of $\rm 1.8\times10^5\,M_\odot$, which is consistent with previous findings \citep[e.g.][]{Immer14} and accounts for $60\%$ of the mass of the hub--filament system. The masses of the filaments hf1, hf2, and hf3 are in the range of $\rm (2.6-4.7)\times10^4\,M_\odot$. These three filaments account for $\sim35\%$ of the mass of the hub--filament system, illustrating that the mass of the hub--filament system is mainly concentrated in the central hub W33.

\subsection{ATLASGAL 870 \um clumps in the hub--filament system}
The ATLASGAL 870 \um emission traces the distribution of cold dust \citep{Beuther12}. From the catalogue of the ATLASGAL 870 \um clumps \citep{Urquhart18}, we extract 49 clumps associated with the hub--filament system based on their $V\rm_{LSR}$ and distances. The physical parameters of these 49 clumps are presented in Table 2. All the clumps are found at a distance of 2.6 kpc, which is consistent with the parallax distance of $\rm 2.4^{+0.17}_{-0.15} \,kpc$ \citep{Immer13}. In addition, the distribution of the clumps with the same distance in the hub--filament system provides  further evidence that the W33 complex and the surrounding filaments constitute a complete system. Also, their $V\rm_{LSR}$ are in the range of $32.7-38.4$ \kms. We overlay the clumps on the C$^{18}$O (1-0) centroid velocity distribution map in Fig. 5a, which reveals that 28 of them are located in the central hub W33. The other clumps are mainly distributed along the spines of the filaments hf3 (10) and hf2 (6).   We also made histograms of radius, dust temperature, mass, and $\rm H_2$ column density of the clumps in the W33 complex and in the filaments; see Fig. 6. We find that the clumps are warmer and denser with larger sizes in the W33 complex, while the mean values of the mass are the same for the clumps in the W33 complex and in the filament.

Furthermore, \citet{Urquhart18} classified $\sim$ 8000 ATLASGAL 870 \um clumps into four evolutionary sequences: quiescent (70 \um weak), protostellar (MIR-dark but FIR-bright), young stellar objects  (YSOs; MIR-bright), and MSF (associated with massive star formation tracers, such as radio-bright HII regions and methanol masers). From column 13 of Table 2, we see that the extracted 49 clumps consist of 12 ($\sim 24\%$) quiescent clumps, 14 ($\sim 29\%$) protostellar clumps, 20 YSO ($\sim 41\%$) clumps, and 3 MSF clumps ($\sim 6\%$). It appears that more than $76\%$ of the clumps are probably forming stars, indicating that the hub--filament system provides a good environment for star formation. In addition, Table 3 presents the percentages of each stage in the central hub W33 and in the filaments. Firstly, we find that only the central hub W33 contains three MSF clumps.  We also find that $70\%$ of the YSO clumps are in the central hub W33, while $\sim 64\%$ of the protostellar clumps are in the filaments. The percentages of the quiescent clumps are equal in the central hub and the filaments.

To investigate the capability of the clumps in the hub--filament system to form massive stars, we consider the relationship between mass and size, as shown in Fig. 7. The yellow shaded region represents a parameter space devoid of massive star formation, as determined by \citet{Kauffmann10}, of $M(r)\geq580{\rm M_\odot}(r/\rm pc)^{1.33}$. From Fig. 7, we find that $\sim 46\%$ of the clumps in the central hub W33 and $\sim 57\%$ of the clumps in the filaments lie above the threshold, suggesting significant potential to form massive stars. Therefore, the hub--filament system is likely to be a suitable environment to form massive stars.

\subsection{Distribution of young stellar objects toward the hub--filament system}
To study the star formation activity in the hub--filament system, we search for YSOs using the GLIMPSE I Spring'07 catalogue. In total, 47787 near-infrared (NIR) sources with 3.6, 4.5, 5.8, and 8.0 \um are selected in the observation zoom. The IRAC [5.8]-[8.0] versus [3.6]-[4.5] colour-colour (CC) diagram is a useful tool for identifying YSOs with infrared excess \citep{Allen04}. Based on the criteria of \citet{Allen04}, these NIR sources are classified into three evolutionary stages, as shown in Fig. 8. Here, 811 NIR sources are identified as Class I sources, which are protostars with circumstellar envelopes and have an age of $\sim 10^5$ yr, and 1252 sources are Class II sources, which are disc-dominated objects with an age of $\sim 10^6$ yr. The remaining NIR sources are other sources (such as classical T Tauri, Herbig Ae/Be). Here, Class I and Class II sources are selected as the YSOs.

As the identified Class II YSOs are almost uniformly distributed in the hub--filament system, we only show the distribution of all the identified Class I sources overlaid on the C$^{18}$O (1-0) emission map in Fig. 9a. We note that Class I sources are preferentially situated in the regions where clumps are clustered in the hub--filament system. The clustering of YSOs in a given area can help us identify the active star-forming regions. To highlight the clustering behaviour of YSOs within the hub--filament system, we analysed the surface density distribution of the identified Class I YSOs in Fig. 9b. The lowest level of the contours is $\sim$ 0.5 pc$^{-2}$ to get rid of the effect of the foreground and background objects. In Fig. 9b, we find that the surface density distribution of the Class I YSOs resembles the structure of the hub--filament system. Moreover, the peaks of the Class I YSO density distribution are located where the clumps group, which appears to suggest that the mechanism of clump and YSO formation in the hub--filament system might be the same.

\section{Discussion}
In order to explore the clump and YSO formation in the hub--filament system, we need to analyse the stability of the central hub W33, which can be calculated using the expression $\alpha_{\rm vir} = 5{\sigma\rm_v}^2 R /GM$, where $R$ is the equivalent radius of the central hub W33, and $ \sigma_{\rm_V} = \sqrt{\sigma_{\rm NT}^2+c\rm_s^2}$ , the mean velocity dispersion. The derived $\alpha\rm_{vir}$ from the optically thin C$^{18}$O (1-0) line is $\sim 0.04$. When $\alpha\rm_{vir}$ is less than 1, the central hub W33 is likely gravitationally bound and perhaps collapsing. At the same time, for the filaments hf1, hf2, and hf3, the difference between the critical mass to length ratio $ (M/L)\rm_{crit}$ and the linear mass $M/L$ can tell us the stability of the filaments \citep{Jackson10}. When $M/L>(M/L)\rm_{crit}$, the filaments are dominated by gravity and may be collapsing. The $(M/L)_{\rm crit}$ can be derived from $ (M/L)_{\rm crit} = 2{\sigma\rm_v}^2/G$, which is mainly caused by turbulence \citep{Jackson10}, and the $M/L$ are calculated using the total masses $M$ and lengths $L$ of the filaments hf1, hf2, and hf3. The derived $M/L$ are the upper limits due to inclination and projection effects. Our computed results are listed in columns 14-15 of Table 1. They indicate that filaments hf1, hf2, and hf3 are probably collapsing globally.

Figure 10 shows the difference between the velocities of the filaments and the junctions as a function of distance to junction. The points in Fig. 10 are extracted along the spine of each filament. From Fig. 10, we can see that the filament hf1 shows a transition on the tail within the last 2 pc, but this might be inaccurate because of possible confusion with other clouds. Indeed, we detect a $\sim 39$ \kms filament there (see Sect.
3.2.1 and Fig. 5b). If ignoring the decreasing trend of hf1, the filaments hf1, hf2, and hf3 all present monotonically increasing profiles. The resultant velocity gradients are consistent with the ones reported in other collapsing filaments \citep{ Peretto14,Liu16,Yuan18}. The velocity gradients for these filaments are estimated to be 0.32, 0.11, and 0.10 km s$^{-1}$ pc$^{-1}$, respectively. The values are comparable to those detected in previous studies \citep{Peretto14,Yuan18}.

Both cloud rotation and accretion flows along the filament can result in this kind of velocity gradient \citep{Veena18}. To examine the rotational feature in the hub--filament system, we constructed position--velocity (PV) diagrams of C$^{18}$O (1-0) along the cuttings in Fig. 3e, as shown in Fig. 11. The white dashed lines mark the location of the W33 complex. The PV plots in Fig. 11 reveal the velocity gradients between the filaments and the W33 complex as well as along the filaments, but no signs of a Keplerian rotation signature are observed. Therefore, we propose that the hub--filament system is not rotating.

On the other hand, accretion flow along the filament could also be causing the smoothed velocity gradient \citep[and reference therein]{Kirk13}. Following \citet{Kirk13}, we derived the gas accretion rate $\dot{M}$ for a simple cylindrical cloud,
\begin{equation}
\dot{M}=\frac{\nabla V M}{\rm tan(\alpha)},
\end{equation}
in which $\alpha$ is assumed as an inclination angle of $45^\circ$ without considering the projection effect \citep{Yuan18, Veena18}. The accretion rates of the filaments are calculated as $8.5\times10^{-3}$, $5.3\times10^{-3}$ , and $3.2\times10^{-3}$ $\rm M_\odot\,yr^{-1}$ for the filaments hf1, hf2, and hf3, respectively, while residual effects from the velocity coherence widely detected in Giant Molecular Filaments (GMFs) cannot be ruled out \citep{Ragan14,Wang15}. From Fig. 5a, we propose that the central hub W33 is probably being fed by the filaments hf1, hf2, and hf3 directly. The total accretion rate is $\rm \sim 1.7\times10^{-2}\, M_\odot\,yr^{-1}$, two orders of magnitude greater than rates found by other studies \citep{Yuan18,Chen19,Trev19}. Given that our velocity gradients are comparable to those in other studies \citep{Yuan18,Chen19,Trev19}, this larger accretion rate could be caused by the two orders of magnitude greater masses of the filaments hf1, hf2, and hf3. Furthermore, this higher accretion rate on the larger scale probably leads to the much larger mass in the W33 complex and probably promotes proto-cluster formation in the W33 complex \citep{Messineo15}. Consequently, the W33 complex is possibly accumulating about $1.7\times10^4$ M$_\odot$/Myr as an upper limit. 

The central hub W33 has a total mass of $\rm\sim 1.8\times10^5\,M_\odot$, accounting for $\sim 60\%$ of the mass of the hub--filament system. This indicates that the central hub is the mass reservoir of the hub--filament system. The clumps in Fig. 5a are clustered in the intersection between filaments hf1, hf2, and hf3 and the central hub W33, indicating that gas flows along the filaments are probably channelling mass to the junction and promoting clump formation in the intersection. Moreover, the other clumps are distributed in the filaments and the W33 complex might be the result of global collapse. Global collapse could be causing the accretion of materials radially into the central zones, providing mass for clump formation.
\section{Summary and conclusions}
We present the large-scale molecular $^{12}$CO (1-0), $^{13}$CO (1-0), and C$^{18}$O (1-0) lines, and infrared observations toward the W33 complex and its surroundings. Our main findings are summarised as follows:

1. According to the centroid velocity distributions of the Gaussian fits to all the $^{13}$CO (1-0) and C$^{18}$O (1-0) lines, six velocity components are distinguished along the line of sight toward the W33 complex from 0 to 70 \kms, namely  2-15 \kms, 15-21 \kms, 21-28 \kms, 30-44 \kms, 44-48 \kms,  and 48-60 \kms. The velocity component 30-44 \kms on large scale is likely to be associated with the W33 complex from the CO emission in Fig. 3d. 

2. The PPV space of C$^{18}$O (1-0) reveals that the 30-44 \kms velocity component consists of three different velocity components in 30-38.5 \kms, 38.5-42 \kms, and 42-44 \kms. The 30-38.5 \kms velocity component shows a real hub--filament system, whose central hub is the W33 complex. Three filaments hf1, hf2, and hf3 are spatially adjacent to the W33 complex in different directions.

3. The central hub W33 has a mass of $\rm 1.8\times10^5\, M_\odot$, accounting for about 60 $\%$ of the mass of the hub--filament system, and the masses of the filaments hf1, hf2, and hf3  are in the range of $(2.6-4.7)\times10^4\, M_\odot$. The mass of the hub--filament system is concentrated in the W33 complex and the adjacent filaments ($\sim 95 \%$).  The central hub W33 is globally collapsing, as suggsted by the virial parameter $\rm \alpha_{vir}$ which is lower than 1. The filaments hf1, hf2, and hf3 are supercritical with mass per unit length ranging from 2214 to 3467 $\rm M_\odot\,pc^{-1}$, indicating that they are also globally collapsing. Regular velocity gradients along the filaments hf1, hf2, and hf3 suggest that these filaments are channelling materials into the junctions  with an accretion rate of $\rm\sim 1.7\times10^{-2}\,M_\odot\, yr^{-1}$.

4. In total, 49 dense clumps are identified to be associated with the hub--filament system, and over 57$\%$ of the clumps (28) are located at the central hub W33, including 3 MSF clumps. The other clumps are situated in the filaments without the MSF clumps. Also, $70\%$ of the YSO clumps of the system are found in the central hub. We find that $\sim 46\%$ of the clumps in the central hub W33 and $\sim 57\%$ of the clumps in the filaments lie above the mass threshold permitting massive star formation, suggesting significant potential to form massive
stars in those regions.

5. The surface density distribution of Class I YSOs in this system shows a similar structure to the hub--filament system, and the peaks in this distribution are at the junctions between the filaments and the W33 complex, where there are concentrations of clumps. Gas flows along the filaments possibly provide the mass for clump formation and probably create the star formation in the intersections.

Based on the above findings, we conclude that the W33 complex is likely to be a central hub of a hub--filament system. Also, the gas flows along the tributary filaments in the system probably promote the proto-cluster formation in the W33 complex.

\begin{acknowledgements}
We are grateful to the staff at the Qinghai Station of PMO for their assistance during the observations. Thanks for the Key Laboratory for Radio Astronomy, CAS, for partly supporting the telescope operation. This work is partly supported by the National Natural Science Foundation of China 11703040 and the National Natural Science Foundation of China 11933011. This work has made used of data from the NASA/IPAC Infrared Science Archive, which is operated by the Jet Propulsion Laboratory, California Institute of Technology, under contract with the National Aeronautics and Space Administration. The ATLASGAL project is a collaboration between the Max-Planck-Gesellschaft, the European Southern Observatory (ESO) and the Universidad de Chile. It includes projects E-181.C-0885, E-078.F-9040(A), M-079.C-9501(A), M-081.C-9501(A) plus Chilean data. C.-P.Z. acknowledges supports from the NAOC Nebula Talents Program, and the Cultivation Project for FAST Scientific Payoff and Research Achievement of CAMS-CAS.
\end{acknowledgements}

\bibliographystyle{bibtex/aa}
\bibliography{bibtex/references.bib}

\begin{thebibliography}{51}
\expandafter\ifx\csname natexlab\endcsname\relax\def\natexlab#1{#1}\fi

\bibitem[{{Allen} {et~al.}(2004){Allen}, {Calvet}, {D'Alessio}, {Merin},
  {Hartmann}, {Megeath}, {Gutermuth}, {Muzerolle}, {Pipher}, {Myers}, \&
  {Fazio}}]{Allen04}
{Allen}, L.~E., {Calvet}, N., {D'Alessio}, P., {et~al.} 2004, \apjs, 154, 363

\bibitem[{{Andr{\'e}} {et~al.}(2014){Andr{\'e}}, {Di Francesco},
  {Ward-Thompson}, {Inutsuka}, {Pudritz}, \& {Pineda}}]{Andre14}
{Andr{\'e}}, P., {Di Francesco}, J., {Ward-Thompson}, D., {et~al.} 2014, in
  Protostars and Planets VI, ed. H.~{Beuther}, R.~S. {Klessen}, C.~P.
  {Dullemond}, \& T.~{Henning}, 27

\bibitem[{{Andr{\'e}} {et~al.}(2010){Andr{\'e}}, {Men'shchikov}, {Bontemps},
  {K{\"o}nyves}, {Motte}, {Schneider}, {Didelon}, {Minier}, {Saraceno},
  {Ward-Thompson}, {di Francesco}, {White}, {Molinari}, {Testi}, {Abergel},
  {Griffin}, {Henning}, {Royer}, {Mer{\'\i}n}, {Vavrek}, {Attard},
  {Arzoumanian}, {Wilson}, {Ade}, {Aussel}, {Baluteau}, {Benedettini},
  {Bernard}, {Blommaert}, {Cambr{\'e}sy}, {Cox}, {di Giorgio}, {Hargrave},
  {Hennemann}, {Huang}, {Kirk}, {Krause}, {Launhardt}, {Leeks}, {Le Pennec},
  {Li}, {Martin}, {Maury}, {Olofsson}, {Omont}, {Peretto}, {Pezzuto}, {Prusti},
  {Roussel}, {Russeil}, {Sauvage}, {Sibthorpe}, {Sicilia-Aguilar}, {Spinoglio},
  {Waelkens}, {Woodcraft}, \& {Zavagno}}]{Andre10}
{Andr{\'e}}, P., {Men'shchikov}, A., {Bontemps}, S., {et~al.} 2010, \aap, 518,
  L102

\bibitem[{{Benjamin} {et~al.}(2003){Benjamin}, {Churchwell}, {Babler}, {Bania},
  {Clemens}, {Cohen}, {Dickey}, {Indebetouw}, {Jackson}, {Kobulnicky},
  {Lazarian}, {Marston}, {Mathis}, {Meade}, {Seager}, {Stolovy}, {Watson},
  {Whitney}, {Wolff}, \& {Wolfire}}]{Benjamin03}
{Benjamin}, R.~A., {Churchwell}, E., {Babler}, B.~L., {et~al.} 2003, \pasp,
  115, 953

\bibitem[{{Beuther} {et~al.}(2012){Beuther}, {Tackenberg}, {Linz}, {Henning},
  {Schuller}, {Wyrowski}, {Schilke}, {Menten}, {Robitaille}, {Walmsley},
  {Bronfman}, {Motte}, {Nguyen-Luong}, \& {Bontemps}}]{Beuther12}
{Beuther}, H., {Tackenberg}, J., {Linz}, H., {et~al.} 2012, \apj, 747, 43

\bibitem[{{Carey} {et~al.}(2009){Carey}, {Noriega-Crespo}, {Mizuno}, {Shenoy},
  {Paladini}, {Kraemer}, {Price}, {Flagey}, {Ryan}, {Ingalls}, {Kuchar},
  {Pinheiro Gon{\c{c}}alves}, {Indebetouw}, {Billot}, {Marleau}, {Padgett},
  {Rebull}, {Bressert}, {Ali}, {Molinari}, {Martin}, {Berriman}, {Boulanger},
  {Latter}, {Miville-Deschenes}, {Shipman}, \& {Testi}}]{Carey09}
{Carey}, S.~J., {Noriega-Crespo}, A., {Mizuno}, D.~R., {et~al.} 2009, \pasp,
  121, 76

\bibitem[{{Chen} {et~al.}(2019){Chen}, {Zhang}, {Wright}, {Busquet}, {Lin},
  {Liu}, {Olguin}, {Sanhueza}, {Nakamura}, {Palau}, {Ohashi}, {Tatematsu}, \&
  {Liao}}]{Chen19}
{Chen}, H.-R.~V., {Zhang}, Q., {Wright}, M.~C.~H., {et~al.} 2019, \apj, 875, 24

\bibitem[{{Clarke} {et~al.}(2018){Clarke}, {Whitworth}, {Spowage},
  {Duarte-Cabral}, {Suri}, {Jaffa}, {Walch}, \& {Clark}}]{Clarke18}
{Clarke}, S.~D., {Whitworth}, A.~P., {Spowage}, R.~L., {et~al.} 2018, \mnras,
  479, 1722

\bibitem[{{Csengeri} {et~al.}(2014){Csengeri}, {Urquhart}, {Schuller}, {Motte},
  {Bontemps}, {Wyrowski}, {Menten}, {Bronfman}, {Beuther}, {Henning}, {Testi},
  {Zavagno}, \& {Walmsley}}]{Csengeri14}
{Csengeri}, T., {Urquhart}, J.~S., {Schuller}, F., {et~al.} 2014, \aap, 565,
  A75

\bibitem[{{Fazio} {et~al.}(2004){Fazio}, {Hora}, {Allen}, {Ashby}, {Barmby},
  {Deutsch}, {Huang}, {Kleiner}, {Marengo}, {Megeath}, {Melnick}, {Pahre},
  {Patten}, {Polizotti}, {Smith}, {Taylor}, {Wang}, {Willner}, {Hoffmann},
  {Pipher}, {Forrest}, {McMurty}, {McCreight}, {McKelvey}, {McMurray}, {Koch},
  {Moseley}, {Arendt}, {Mentzell}, {Marx}, {Losch}, {Mayman}, {Eichhorn},
  {Krebs}, {Jhabvala}, {Gezari}, {Fixsen}, {Flores}, {Shakoorzadeh}, {Jungo},
  {Hakun}, {Workman}, {Karpati}, {Kichak}, {Whitley}, {Mann}, {Tollestrup},
  {Eisenhardt}, {Stern}, {Gorjian}, {Bhattacharya}, {Carey}, {Nelson},
  {Glaccum}, {Lacy}, {Lowrance}, {Laine}, {Reach}, {Stauffer}, {Surace},
  {Wilson}, {Wright}, {Hoffman}, {Domingo}, \& {Cohen}}]{Fazio04}
{Fazio}, G.~G., {Hora}, J.~L., {Allen}, L.~E., {et~al.} 2004, \apjs, 154, 10

\bibitem[{{Frerking} {et~al.}(1982){Frerking}, {Langer}, \&
  {Wilson}}]{Frerking82}
{Frerking}, M.~A., {Langer}, W.~D., \& {Wilson}, R.~W. 1982, \apj, 262, 590

\bibitem[{{Friesen} {et~al.}(2016){Friesen}, {Bourke}, {Di Francesco},
  {Gutermuth}, \& {Myers}}]{Friesen16}
{Friesen}, R.~K., {Bourke}, T.~L., {Di Francesco}, J., {Gutermuth}, R., \&
  {Myers}, P.~C. 2016, \apj, 833, 204

\bibitem[{{Garden91} {et~al.}(1991){Garden91}, {Hayashi}, {Gatley}, {Hasegawa},
  \& {Kaifu}}]{Garden91}
{Garden91}, R.~P., {Hayashi}, M., {Gatley}, I., {Hasegawa}, T., \& {Kaifu}, N.
  1991, \apj, 374, 540

\bibitem[{{Gardner} \& {Whiteoak}(1972)}]{Gardner72}
{Gardner}, F.~F. \& {Whiteoak}, J.~B. 1972, \aplett, 12, 107

\bibitem[{{Goldsmith} \& {Mao}(1983)}]{Goldsmith83}
{Goldsmith}, P.~F. \& {Mao}, X.~J. 1983, \apj, 265, 791

\bibitem[{{G{\'o}mez} \& {V{\'a}zquez-Semadeni}(2014)}]{Gomez14}
{G{\'o}mez}, G.~C. \& {V{\'a}zquez-Semadeni}, E. 2014, \apj, 791, 124

\bibitem[{{G{\"u}sten} {et~al.}(2006{\natexlab{a}}){G{\"u}sten}, {Booth},
  {Cesarsky}, {Menten}, {Agurto}, {Anciaux}, {Azagra}, {Belitsky}, {Belloche},
  {Bergman}, {De Breuck}, {Comito}, {Dumke}, {Duran}, {Esch}, {Fluxa}, {Greve},
  {Hafok}, {H{\"a}upl}, {Helldner}, {Henseler}, {Heyminck}, {Johansson},
  {Kasemann}, {Klein}, {Korn}, {Kreysa}, {Kurz}, {Lapkin}, {Leurini}, {Lis},
  {Lundgren}, {Mac-Auliffe}, {Martinez}, {Melnick}, {Morris}, {Muders},
  {Nyman}, {Olberg}, {Olivares}, {Pantaleev}, {Patel}, {Pausch}, {Philipp},
  {Philipps}, {Sridharan}, {Polehampton}, {Reveret}, {Risacher}, {Roa},
  {Sauer}, {Schilke}, {Santana}, {Schneider}, {Sepulveda}, {Siringo},
  {Spyromilio}, {Stenvers}, {van der Tak}, {Torres}, {Vanzi}, {Vassilev},
  {Weiss}, {Willmeroth}, {Wunsch}, \& {Wyrowski}}]{Gusten06a}
{G{\"u}sten}, R., {Booth}, R.~S., {Cesarsky}, C., {et~al.} 2006{\natexlab{a}},
  in Society of Photo-Optical Instrumentation Engineers (SPIE) Conference
  Series, Vol. 6267, Society of Photo-Optical Instrumentation Engineers (SPIE)
  Conference Series, ed. L.~M. {Stepp}, 626714

\bibitem[{{G{\"u}sten} {et~al.}(2006{\natexlab{b}}){G{\"u}sten}, {Nyman},
  {Schilke}, {Menten}, {Cesarsky}, \& {Booth}}]{Gusten06b}
{G{\"u}sten}, R., {Nyman}, L.~{\r{A}}., {Schilke}, P., {et~al.}
  2006{\natexlab{b}}, \aap, 454, L13

\bibitem[{{Guzm{\'a}n} {et~al.}(2015){Guzm{\'a}n}, {Sanhueza}, {Contreras},
  {Smith}, {Jackson}, {Hoq}, \& {Rathborne}}]{Guzman15}
{Guzm{\'a}n}, A.~E., {Sanhueza}, P., {Contreras}, Y., {et~al.} 2015, \apj, 815,
  130

\bibitem[{{Hennemann} {et~al.}(2014){Hennemann}, {Motte}, \&
  {Schneider}}]{Hennemann14}
{Hennemann}, M., {Motte}, F., \& {Schneider}, N. 2014, in The Labyrinth of Star
  Formation, Vol.~36, 271

\bibitem[{{Immer} {et~al.}(2014){Immer}, {Galv{\'a}n-Madrid}, {K{\"o}nig},
  {Liu}, \& {Menten}}]{Immer14}
{Immer}, K., {Galv{\'a}n-Madrid}, R., {K{\"o}nig}, C., {Liu}, H.~B., \&
  {Menten}, K.~M. 2014, \aap, 572, A63

\bibitem[{{Immer} {et~al.}(2013){Immer}, {Reid}, {Menten}, {Brunthaler}, \&
  {Dame}}]{Immer13}
{Immer}, K., {Reid}, M.~J., {Menten}, K.~M., {Brunthaler}, A., \& {Dame}, T.~M.
  2013, \aap, 553, A117

\bibitem[{{Jackson} {et~al.}(2010){Jackson}, {Finn}, {Chambers}, {Rathborne},
  \& {Simon}}]{Jackson10}
{Jackson}, J.~M., {Finn}, S.~C., {Chambers}, E.~T., {Rathborne}, J.~M., \&
  {Simon}, R. 2010, \apjl, 719, L185

\bibitem[{{Kauffmann} {et~al.}(2008){Kauffmann}, {Bertoldi}, {Bourke}, {Evans},
  \& {Lee}}]{Kauffmann08}
{Kauffmann}, J., {Bertoldi}, F., {Bourke}, T.~L., {Evans}, N.~J., I., \& {Lee},
  C.~W. 2008, \aap, 487, 993

\bibitem[{{Kauffmann} \& {Pillai}(2010)}]{Kauffmann10}
{Kauffmann}, J. \& {Pillai}, T. 2010, \apjl, 723, L7

\bibitem[{{Kirk} {et~al.}(2013){Kirk}, {Myers}, {Bourke}, {Gutermuth},
  {Hedden}, \& {Wilson}}]{Kirk13}
{Kirk}, H., {Myers}, P.~C., {Bourke}, T.~L., {et~al.} 2013, \apj, 766, 115

\bibitem[{{Kohno} {et~al.}(2018){Kohno}, {Torii}, {Tachihara}, {Umemoto},
  {Minamidani}, {Nishimura}, {Fujita}, {Matsuo}, {Yamagishi}, {Tsuda},
  {Kuriki}, {Kuno}, {Ohama}, {Hattori}, {Sano}, {Yamamoto}, \&
  {Fukui}}]{Kohno18}
{Kohno}, M., {Torii}, K., {Tachihara}, K., {et~al.} 2018, \pasj, 70, S50

\bibitem[{{Kumar} {et~al.}(2020){Kumar}, {Palmeirim}, {Arzoumanian}, \&
  {Inutsuka}}]{Kumar2020}
{Kumar}, M.~S.~N., {Palmeirim}, P., {Arzoumanian}, D., \& {Inutsuka}, S.~I.
  2020, \aap, 642, A87

\bibitem[{{Liu} {et~al.}(2012){Liu}, {Hsieh}, {Ho}, {Su}, {Wright}, {Sun}, \&
  {Minh}}]{Liu12}
{Liu}, H.~B., {Hsieh}, P.-Y., {Ho}, P. T.~P., {et~al.} 2012, \apj, 756, 195

\bibitem[{{Liu} {et~al.}(2016){Liu}, {Zhang}, {Kim}, {Wu}, {Lee}, {Goldsmith},
  {Li}, {Liu}, {Chen}, {Tatematsu}, {Wang}, {Lee}, {Qin}, {Mardones}, \&
  {Cho}}]{Liu16}
{Liu}, T., {Zhang}, Q., {Kim}, K.-T., {et~al.} 2016, \apj, 824, 31

\bibitem[{{Liu} {et~al.}(2018){Liu}, {Xu}, {Ning}, {Zhang}, \& {Liu}}]{Liu18}
{Liu}, X.-L., {Xu}, J.-L., {Ning}, C.-C., {Zhang}, C.-P., \& {Liu}, X.-T. 2018,
  Research in Astronomy and Astrophysics, 18, 004

\bibitem[{{Messineo} {et~al.}(2015){Messineo}, {Clark}, {Figer}, {Kudritzki},
  {Najarro}, {Rich}, {Menten}, {Ivanov}, {Valenti}, {Trombley}, {Chen}, \&
  {Davies}}]{Messineo15}
{Messineo}, M., {Clark}, J.~S., {Figer}, D.~F., {et~al.} 2015, \apj, 805, 110

\bibitem[{{Morgan} {et~al.}(2010){Morgan}, {Figura}, {Urquhart}, \&
  {Thompson}}]{Morgan10}
{Morgan}, L.~K., {Figura}, C.~C., {Urquhart}, J.~S., \& {Thompson}, M.~A. 2010,
  \mnras, 408, 157

\bibitem[{{M{\"u}ller} {et~al.}(2005){M{\"u}ller}, {Schl{\"o}der}, {Stutzki},
  \& {Winnewisser}}]{Muller05}
{M{\"u}ller}, H. S.~P., {Schl{\"o}der}, F., {Stutzki}, J., \& {Winnewisser}, G.
  2005, Journal of Molecular Structure, 742, 215

\bibitem[{{M{\"u}ller} {et~al.}(2001){M{\"u}ller}, {Thorwirth}, {Roth}, \&
  {Winnewisser}}]{Muller01}
{M{\"u}ller}, H.~S.~P., {Thorwirth}, S., {Roth}, D.~A., \& {Winnewisser}, G.
  2001, \aap, 370, L49

\bibitem[{{Myers}(2009)}]{Myers09}
{Myers}, P.~C. 2009, \apj, 700, 1609

\bibitem[{{Peretto} {et~al.}(2014){Peretto}, {Fuller}, {Andr{\'e}},
  {Arzoumanian}, {Rivilla}, {Bardeau}, {Duarte Puertas}, {Guzman Fernandez},
  {Lenfestey}, {Li}, {Olguin}, {R{\"o}ck}, {de Villiers}, \&
  {Williams}}]{Peretto14}
{Peretto}, N., {Fuller}, G.~A., {Andr{\'e}}, P., {et~al.} 2014, \aap, 561, A83

\bibitem[{{Peretto} {et~al.}(2013){Peretto}, {Fuller}, {Duarte-Cabral},
  {Avison}, {Hennebelle}, {Pineda}, {Andr{\'e}}, {Bontemps}, {Motte},
  {Schneider}, \& {Molinari}}]{Peretto13}
{Peretto}, N., {Fuller}, G.~A., {Duarte-Cabral}, A., {et~al.} 2013, \aap, 555,
  A112

\bibitem[{{Peretto} {et~al.}(2016){Peretto}, {Lenfestey}, {Fuller},
  {Traficante}, {Molinari}, {Thompson}, \& {Ward-Thompson}}]{Peretto16}
{Peretto}, N., {Lenfestey}, C., {Fuller}, G.~A., {et~al.} 2016, \aap, 590, A72

\bibitem[{{Pety}(2005)}]{Pety05}
{Pety}, J. 2005, in SF2A-2005: Semaine de l'Astrophysique Francaise, ed.
  F.~{Casoli}, T.~{Contini}, J.~M. {Hameury}, \& L.~{Pagani}, 721

\bibitem[{{Pilbratt} {et~al.}(2010){Pilbratt}, {Riedinger}, {Passvogel},
  {Crone}, {Doyle}, {Gageur}, {Heras}, {Jewell}, {Metcalfe}, {Ott}, \&
  {Schmidt}}]{Pilbratt10}
{Pilbratt}, G.~L., {Riedinger}, J.~R., {Passvogel}, T., {et~al.} 2010, \aap,
  518, L1

\bibitem[{{Pineda} {et~al.}(2008){Pineda}, {Caselli}, \& {Goodman}}]{Pineda08}
{Pineda}, J.~E., {Caselli}, P., \& {Goodman}, A.~A. 2008, \apj, 679, 481

\bibitem[{{Ragan} {et~al.}(2014){Ragan}, {Henning}, {Tackenberg}, {Beuther},
  {Johnston}, {Kainulainen}, \& {Linz}}]{Ragan14}
{Ragan}, S.~E., {Henning}, T., {Tackenberg}, J., {et~al.} 2014, \aap, 568, A73

\bibitem[{{Sanhueza} {et~al.}(2012){Sanhueza}, {Jackson}, {Foster}, {Garay},
  {Silva}, \& {Finn}}]{Sanhueza12}
{Sanhueza}, P., {Jackson}, J.~M., {Foster}, J.~B., {et~al.} 2012, \apj, 756, 60

\bibitem[{{Schneider} {et~al.}(2012){Schneider}, {Csengeri}, {Hennemann},
  {Motte}, {Didelon}, {Federrath}, {Bontemps}, {Di Francesco}, {Arzoumanian},
  {Minier}, {Andr{\'e}}, {Hill}, {Zavagno}, {Nguyen-Luong}, {Attard},
  {Bernard}, {Elia}, {Fallscheer}, {Griffin}, {Kirk}, {Klessen}, {K{\"o}nyves},
  {Martin}, {Men'shchikov}, {Palmeirim}, {Peretto}, {Pestalozzi}, {Russeil},
  {Sadavoy}, {Sousbie}, {Testi}, {Tremblin}, {Ward-Thompson}, \&
  {White}}]{Schneider12}
{Schneider}, N., {Csengeri}, T., {Hennemann}, M., {et~al.} 2012, \aap, 540, L11

\bibitem[{{Schuller} {et~al.}(2009){Schuller}, {Menten}, {Contreras},
  {Wyrowski}, {Schilke}, {Bronfman}, {Henning}, {Walmsley}, {Beuther},
  {Bontemps}, {Cesaroni}, {Deharveng}, {Garay}, {Herpin}, {Lefloch}, {Linz},
  {Mardones}, {Minier}, {Molinari}, {Motte}, {Nyman}, {Reveret}, {Risacher},
  {Russeil}, {Schneider}, {Testi}, {Troost}, {Vasyunina}, {Wienen}, {Zavagno},
  {Kovacs}, {Kreysa}, {Siringo}, \& {Wei{\ss}}}]{Schuller09}
{Schuller}, F., {Menten}, K.~M., {Contreras}, Y., {et~al.} 2009, \aap, 504, 415

\bibitem[{{Trevi{\~n}o-Morales} {et~al.}(2019){Trevi{\~n}o-Morales}, {Fuente},
  {S{\'a}nchez-Monge}, {Kainulainen}, {Didelon}, {Suri}, {Schneider},
  {Ballesteros-Paredes}, {Lee}, {Hennebelle}, {Pilleri},
  {Gonz{\'a}lez-Garc{\'\i}a}, {Kramer}, {Garc{\'\i}a-Burillo}, {Luna},
  {Goicoechea}, {Tremblin}, \& {Geen}}]{Trev19}
{Trevi{\~n}o-Morales}, S.~P., {Fuente}, A., {S{\'a}nchez-Monge}, {\'A}.,
  {et~al.} 2019, \aap, 629, A81

\bibitem[{{Urquhart} {et~al.}(2018){Urquhart}, {K{\"o}nig}, {Giannetti},
  {Leurini}, {Moore}, {Eden}, {Pillai}, {Thompson}, {Braiding}, {Burton},
  {Csengeri}, {Dempsey}, {Figura}, {Froebrich}, {Menten}, {Schuller}, {Smith},
  \& {Wyrowski}}]{Urquhart18}
{Urquhart}, J.~S., {K{\"o}nig}, C., {Giannetti}, A., {et~al.} 2018, \mnras,
  473, 1059

\bibitem[{{Veena} {et~al.}(2018){Veena}, {Vig}, {Mookerjea},
  {S{\'a}nchez-Monge}, {Tej}, \& {Ishwara-Chandra}}]{Veena18}
{Veena}, V.~S., {Vig}, S., {Mookerjea}, B., {et~al.} 2018, \apj, 852, 93

\bibitem[{{Wang} {et~al.}(2015){Wang}, {Testi}, {Ginsburg}, {Walmsley},
  {Molinari}, \& {Schisano}}]{Wang15}
{Wang}, K., {Testi}, L., {Ginsburg}, A., {et~al.} 2015, \mnras, 450, 4043

\bibitem[{{Yuan} {et~al.}(2018){Yuan}, {Li}, {Wu}, {Ellingsen}, {Henkel},
  {Wang}, {Liu}, {Liu}, {Zavagno}, {Ren}, \& {Huang}}]{Yuan18}
{Yuan}, J., {Li}, J.-Z., {Wu}, Y., {et~al.} 2018, \apj, 852, 12

\end{thebibliography}

\clearpage
\onecolumn

\begin{landscape}
\setlength{\LTleft}{0pt} \setlength{\LTright}{0pt}
{\tiny
 \renewcommand{\footnoterule}{}
  \begin{longtable}{cccccccccccccc}
   \caption{\label{tab1} Main properties of the filaments and the W33 complex in the hub-filament system. }\\
      \hline
Name & $<l>$  & $<b>$   & Velocity range   & $<T\rm _{ex}>$ & $<\sigma>$ & $<\sigma_{\rm NT}/c\rm_s>$  & $\rm<log(N\rm (H_2))>$  & $\rm<log(n\rm (H_2))>$  & Mass  &Length & Width & $M/L$ & $ (M/L)_{\rm crit}$
\\
  &[deg]  &[deg]  &[\kms]  &[K]  &[\kms]   &    &[$\rm cm^{-2}$]   &[$\rm cm^{-3}$] &$\rm M_\odot$  &[pc] &[pc] &[$\rm M_\odot\,pc^{-1}$]  &[$\rm M_\odot\,pc^{-1}$] \\
(1)   &(2)    &(3)  &(4)    &(5)  &(6) &(7) &(8) &(9) &(10) &(11) &(12) &(13) &(14) \\
\hline
\endfirsthead
\caption{continued.}\\
\hline
Name  & $<l>$  & $<b>$  & Velocity range  & $<T\rm _{ex}>$ & $<\sigma>$ & $<\sigma_{\rm NT}/c\rm_s>$  & $\rm<log(N\rm (H_2))>$  & $\rm<log(n\rm (H_2))>$  & $\rm log(Mass)$ &Length & Width  & $M/L$ & $ (M/L)_{\rm crit}$
\\
 &[deg]  &[deg]  &[\kms]  &[\kms]  &[K]   &   &[$\rm cm^{-2}$]   &[$\rm cm^{-3}$] &$\rm M_\odot$  &[pc] &[pc] &[$\rm M_\odot\,pc^{-1}$]  &[$\rm M_\odot\,pc^{-1}$] \\
(1)   &(2)    &(3)  &(4)    &(5)  &(6) &(7) &(8) &(9) &(10) &(11) &(12) &(13) &(14) \\
\hline
\endhead
\hline
\endfoot
the hub-filament system & 12.951 & -0.247 & 30.20 - 38.50 & 17.49 &  0.90 &  1.36 & 22.43 &  4.39 & $3.0\times10^5$  & ---  & --- & ---    & ---                  \\
 the W33 complex & 12.849 & -0.247 & 30.20 - 38.50 & 20.74 &  0.98 &  1.36 & 22.48 &  4.45 & $1.8\times10^5$         & 7.3  & 5.0 & ---    & ---                   \\
             hf1 & 12.869 & -0.387 & 32.56 - 38.50 & 14.45 &  0.94 &  2.02 & 22.29 &  4.26 & $2.6\times10^4$         & 7.5  & 4.6 &3467    &429           \\
             hf2 & 13.090 & -0.320 & 31.72 - 38.50 & 14.78 &  0.90 &  1.98 & 22.29 &  4.25 & $4.7\times10^4$         & 15.7 & 3.8 &2994    &395            \\
             hf3 & 13.119 & -0.133 & 32.47 - 38.43 & 16.05 &  0.80 &  2.06 & 22.25 &  4.22 & $3.1\times10^4$         & 14.0 & 3.0 &2214    &318           \\

      \hline
\end{longtable}
\small{Notes. Column 1: Central hub W33 and the filaments shown in Fig. 5a.  Columns 2-3: Mean coordinates. Column 4: Spanned velocity ranges. Column 5: Mean excitation temperatures. Column 6: Mean velocity dispersions. Column 7: Mean non-thermal velocity dispersions in sound speed units. Column 8: Mean H$_2$ column densities. Column 9: Mean H$_2$ number densities. Column 10: Total mass. Column 11: Lengths of the filaments. Column 12: Widths of the filaments. Column 13: Linear mass. Column 14: Critical mass to length ratios.}\\
}
\end{landscape}

\clearpage
\begin{landscape}
\setlength{\LTleft}{0pt} \setlength{\LTright}{0pt}
{\tiny
 \renewcommand{\footnoterule}{}
  \begin{longtable}{ccccccccccccccc}
   \caption{\label{tab2} Summary of the parameters for the 49 ATLASGAL clumps in the hub-filament system. } \\
      \hline
Name    &l &b &$V\rm_{ref}$ & distance  & radius  & $S\rm_{peak}$  &$S\rm_{int}$  &$T\rm_{dust}$  &$\rm log(Mass)$  &$\rm log(N\rm(H_2))$     &classification \\

   &[deg] &[deg] &[\kms]  & [kpc]  & [pc]  &[Jy beam$^{-1}$]  &[Jy]  &[K]  &[$\rm M_\odot$] &[$\rm cm^{-2}$]     &  \\
(1)   &(2)    &(3)            &(4)    &(5)  &(6) &(7) &(8) &(9) &(10) &(11) &(12) \\
\hline
\endfirsthead
\caption{continued.}\\
\hline
Name    &l &b &$V\rm_{ref}$ & distance  & radius  & $S\rm_{peak}$  &$S\rm_{int}$  &$T\rm_{dust}$  &$\rm log(Mass)$  &$\rm log(N\rm(H_2))$     &classification \\

   &[deg] &[deg] &[\kms]  & [kpc]  & [pc]  &[Jy beam$^{-1}$]  &[Jy]  &[K]  &[$\rm M_\odot$] &[$\rm cm^{-2}$]     &  \\
(1)   &(2)    &(3)            &(4)    &(5)  &(6) &(7) &(8) &(9) &(10) &(11) &(12) \\
\hline
\endhead
\hline
\endfoot
AGAL012.554-00.347  & 12.554  & -0.347   &  35.7    &  2.6   &  0.57  &   0.53 &    3.74  &  14.00     &   2.38 &   22.37   &    Protostellar  \\
AGAL012.614+00.244  & 12.614  &  0.244   &  33.0    &  2.6   &  0.30  &   0.38 &    3.05  &  15.10     &   2.23 &   22.17   &    Protostellar  \\
AGAL012.648-00.167  & 12.648  & -0.167   &  34.6    &  2.6   &  0.30  &   0.66 &    2.51  &  14.20     &   2.19 &   22.45   &    Protostellar  \\
AGAL012.661-00.166  & 12.661  & -0.166   &  34.0    &  2.6   &  0.79  &   0.89 &    8.17  &  17.90     &   2.55 &   22.42   &             YSO  \\
AGAL012.691-00.157  & 12.691  & -0.157   &  33.1    &  2.6   &  0.33  &   0.76 &    4.34  &  17.50     &   2.29 &   22.37   &    Protostellar  \\
AGAL012.694-00.129  & 12.694  & -0.129   &  32.8    &  2.6   &  0.30  &   0.50 &    1.82  &  16.60     &   1.95 &   22.22   &             YSO  \\
AGAL012.703-00.144  & 12.703  & -0.144   &  34.7    &  2.6   &  0.94  &   0.74 &    8.69  &  22.00     &   2.44 &   22.21   &             YSO  \\
AGAL012.719-00.217  & 12.719  & -0.217   &  34.1    &  2.6   &  1.48  &   3.09 &   21.09  &  25.40     &   2.74 &   22.74   &             YSO  \\
AGAL012.748-00.226  & 12.748  & -0.226   &  36.1    &  2.6   &  0.30  &   0.49 &    3.38  &  24.50     &   1.97 &   21.96   &       Quiescent  \\
AGAL012.776-00.211  & 12.776  & -0.211   &  35.0    &  2.6   &  1.03  &   1.93 &   14.73  &  21.10     &   2.70 &   22.65   &    Protostellar  \\
AGAL012.804-00.199  & 12.804  & -0.199   &  35.4    &  2.6   &  3.09  &  36.48 &  339.23  &  35.10$\rm^a$     &   3.85$\rm^a$ &   23.71$\rm^a$   &             MSF  \\
AGAL012.831-00.207  & 12.831  & -0.207   &  32.5    &  2.6   &  0.30  &   3.19 &    6.50  &  22.80     &   2.29 &   22.82   &             YSO  \\
AGAL012.838-00.286  & 12.838  & -0.286   &  33.0    &  2.6   &  0.48  &   0.90 &    2.65  &  11.90     &   2.35 &   22.72   &             YSO  \\
AGAL012.844-00.352  & 12.844  & -0.352   &  36.5    &  2.6   &  1.00  &   0.50 &    7.63  &  14.30     &   2.67 &   22.33   &    Protostellar  \\
AGAL012.853-00.226  & 12.853  & -0.226   &  36.3    &  2.6   &  1.75  &   3.35 &   26.59  &  21.70     &   2.94 &   22.87   &             YSO  \\
AGAL012.856-00.209  & 12.856  & -0.209   &  32.7    &  2.6   &  0.85  &   1.14 &    7.84  &  18.30     &   2.51 &   22.51   &       Quiescent  \\
AGAL012.858-00.272  & 12.858  & -0.272   &  36.7    &  2.6   &  1.21  &   3.16 &   13.65  &  26.60     &   2.52 &   22.73   &             MSF  \\
AGAL012.861-00.336  & 12.861  & -0.336   &  32.6    &  2.6   &  0.42  &   0.37 &    3.78  &  13.00     &   2.44 &   22.27   &       Quiescent  \\
AGAL012.878-00.287  & 12.878  & -0.287   &  32.8    &  2.6   &  0.91  &   1.84 &   10.16  &  23.40     &   2.47 &   22.57   &             YSO  \\
AGAL012.879-00.316  & 12.879  & -0.316   &  37.6    &  2.6   &  0.30  &   0.48 &    2.48  &  15.20     &   2.14 &   22.27   &    Protostellar  \\
AGAL012.883-00.254  & 12.883  & -0.254   &  35.6    &  2.6   &  0.82  &   1.18 &    7.82  &  17.60$\rm^a$     &   2.55$\rm^a$ &   22.57$\rm^a$   &             YSO  \\
AGAL012.893-00.282  & 12.893  & -0.282   &  34.5    &  2.6   &  1.27  &   1.94 &   21.12  &  18.80     &   2.93 &   22.73   &             YSO  \\
AGAL012.899-00.241  & 12.899  & -0.241   &  36.5    &  2.6   &  0.67  &   3.06 &   12.92  &  17.60     &   2.76 &   22.97   &    Protostellar  \\
AGAL012.908-00.259  & 12.908  & -0.259   &  36.7    &  2.6   &  2.48  &   9.19 &  100.07  &  23.60     &   3.46 &   23.26   &             MSF  \\
AGAL012.914-00.336  & 12.914  & -0.336   &  37.0    &  2.6   &  0.85  &   1.36 &    5.37  &  14.20     &   2.53 &   22.77   &             YSO  \\
AGAL012.929-00.319  & 12.929  & -0.319   &  35.3    &  2.6   &  0.30  &   0.48 &    1.11  &  13.50     &   1.88 &   22.35   &       Quiescent  \\
AGAL012.931-00.074  & 12.931  & -0.074   &  30.6    &  2.6   &  0.33  &   0.67 &    2.47  &  22.70     &   1.88 &   22.15   &             YSO  \\
AGAL012.934-00.274  & 12.934  & -0.274   &  36.0    &  2.6   &  0.30  &   1.07 &    4.02  &  21.30     &   2.13 &   22.39   &             YSO  \\
AGAL012.943-00.119  & 12.943  & -0.119   &  36.2    &  2.6   &  0.24  &   0.58 &    3.27  &  12.10     &   2.43 &   22.52   &       Quiescent  \\
AGAL012.943-00.234  & 12.943  & -0.234   &  33.3    &  2.6   &  0.45  &   0.87 &    7.52  &  15.20     &   2.62 &   22.52   &             YSO  \\
AGAL012.946-00.249  & 12.946  & -0.249   &  34.0    &  2.6   &  0.88  &   0.99 &    7.62  &   9.60     &   3.00 &   22.95   &       Quiescent  \\
AGAL012.954-00.229  & 12.954  & -0.229   &  33.4    &  2.6   &  2.33  &   1.09 &   15.35  &  15.70     &   2.91 &   22.60   &             YSO  \\
AGAL012.968-00.237  & 12.968  & -0.237   &  33.7    &  2.6   &  1.00  &   0.83 &    6.39  &  13.70     &   2.63 &   22.58   &       Quiescent  \\
AGAL013.016-00.179  & 13.016  & -0.179   &  35.1    &  2.6   &  1.33  &   0.66 &   14.81  &  14.10     &   2.97 &   22.46   &    Protostellar  \\
AGAL013.034-00.336  & 13.034  & -0.336   &  37.9    &  2.6   &  0.51  &   0.53 &    3.08  &  13.10     &   2.35 &   22.42   &       Quiescent  \\
AGAL013.066-00.172  & 13.066  & -0.172   &  35.8    &  2.6   &  1.36  &   0.56 &   15.21  &  14.70     &   2.95 &   22.36   &       Quiescent  \\
AGAL013.074-00.316  & 13.074  & -0.316   &  38.0    &  2.6   &  0.30  &   0.40 &    4.86  &  26.00     &   2.09 &   21.84   &             YSO  \\
AGAL013.078-00.334  & 13.078  & -0.334   &  36.8    &  2.6   &  0.30  &   0.47 &    3.13  &  17.20     &   2.16 &   22.17   &             YSO  \\
AGAL013.119-00.096  & 13.119  & -0.096   &  36.4    &  2.6   &  1.60  &   0.95 &   10.36  &  12.60     &   2.90 &   22.70   &    Protostellar  \\
AGAL013.134-00.081  & 13.134  & -0.081   &  36.8    &  2.6   &  0.51  &   0.68 &    2.31  &  15.40     &   2.10 &   22.41   &    Protostellar  \\
AGAL013.158-00.099  & 13.158  & -0.099   &  38.0    &  2.6   &  0.30  &   0.60 &    3.21  &  12.20     &   2.42 &   22.53   &             YSO  \\
AGAL013.164-00.081  & 13.164  & -0.081   &  36.8    &  2.6   &  1.97  &   0.93 &   14.77  &  16.60     &   2.85 &   22.49   &             YSO  \\
AGAL013.171-00.341  & 13.171  & -0.341   &  36.4    &  2.6   &  1.09  &   0.47 &    6.00  &  11.70     &   2.72 &   22.45   &       Quiescent  \\
AGAL013.213-00.077  & 13.213  & -0.077   &  36.6    &  2.6   &  0.30  &   0.53 &    2.90  &   9.60     &   2.58 &   22.68   &       Quiescent  \\
AGAL013.214-00.011  & 13.214  & -0.011   &  34.1    &  2.6   &  0.30  &   0.42 &    2.43  &  20.00     &   1.95 &   22.02   &             YSO  \\
AGAL013.234-00.071  & 13.234  & -0.071   &  34.6    &  2.6   &  0.54  &   0.51 &    4.02  &  12.90     &   2.47 &   22.41   &    Protostellar  \\
AGAL013.243-00.086  & 13.243  & -0.086   &  36.4    &  2.6   &  1.30  &   2.49 &   14.07  &  12.70     &   3.03 &   23.11   &    Protostellar  \\
AGAL013.299-00.402  & 13.299  & -0.402   &  37.2    &  2.6   &  0.30  &   0.41 &    1.90  &  12.70     &   2.16 &   22.33   &    Protostellar  \\
AGAL013.303-00.442  & 13.303  & -0.442   &  33.6    &  2.6   &  0.61  &   0.55 &    4.96  &  14.10     &   2.50 &   22.38   &       Quiescent  \\

      \hline
\end{longtable}
Notes: Column 1: Name of the clumps. Columns 2-3: Peak coordinates. Column 4:  LSR velocities from \citet{Urquhart18}. Column 5: Distances. Column 6: Equivalent radii. Column 7: Peak fluxes at 870 \um. Column 8: Integrated fluxes at 870 \um. Column 9: Dust temperatures. Column 10: Masses. Column 11:  H$_2$ column densities. Column 12: Classification in different evolutional stages.
$^{\rm a}$  values from \citet{Guzman15} \\
}
\end{landscape}

\setlength{\LTleft}{0pt} \setlength{\LTright}{0pt}
\begin{table*}
\begin{center}
   \tabcolsep 4.5mm\caption{Percentages of the clumps in each stage in different places. }
   \label{tab3}
   \small
\def\temptablewidth{10\textwidth}%
    \begin{tabular}{ccccccccccr}
     \hline\noalign{\smallskip}
Name  & Quiescent clumps   & Protostellar clumps   & YSO clumps  &MSF clumps  \\ \hline

The W33 complex   & $50\%$(6)    &$36\%$(5)     &$70\%$(14)      &$100\%$(3)  \\
Filaments         & $50\%$(6)    &$64\%$(9)     &$30\%$(6)       &0(0))        \\

      \noalign{\smallskip}\hline
      \end{tabular}
      \end{center}
      \end{table*}

\begin{figure*}
  \centering
   \includegraphics[angle=0,scale=0.5]{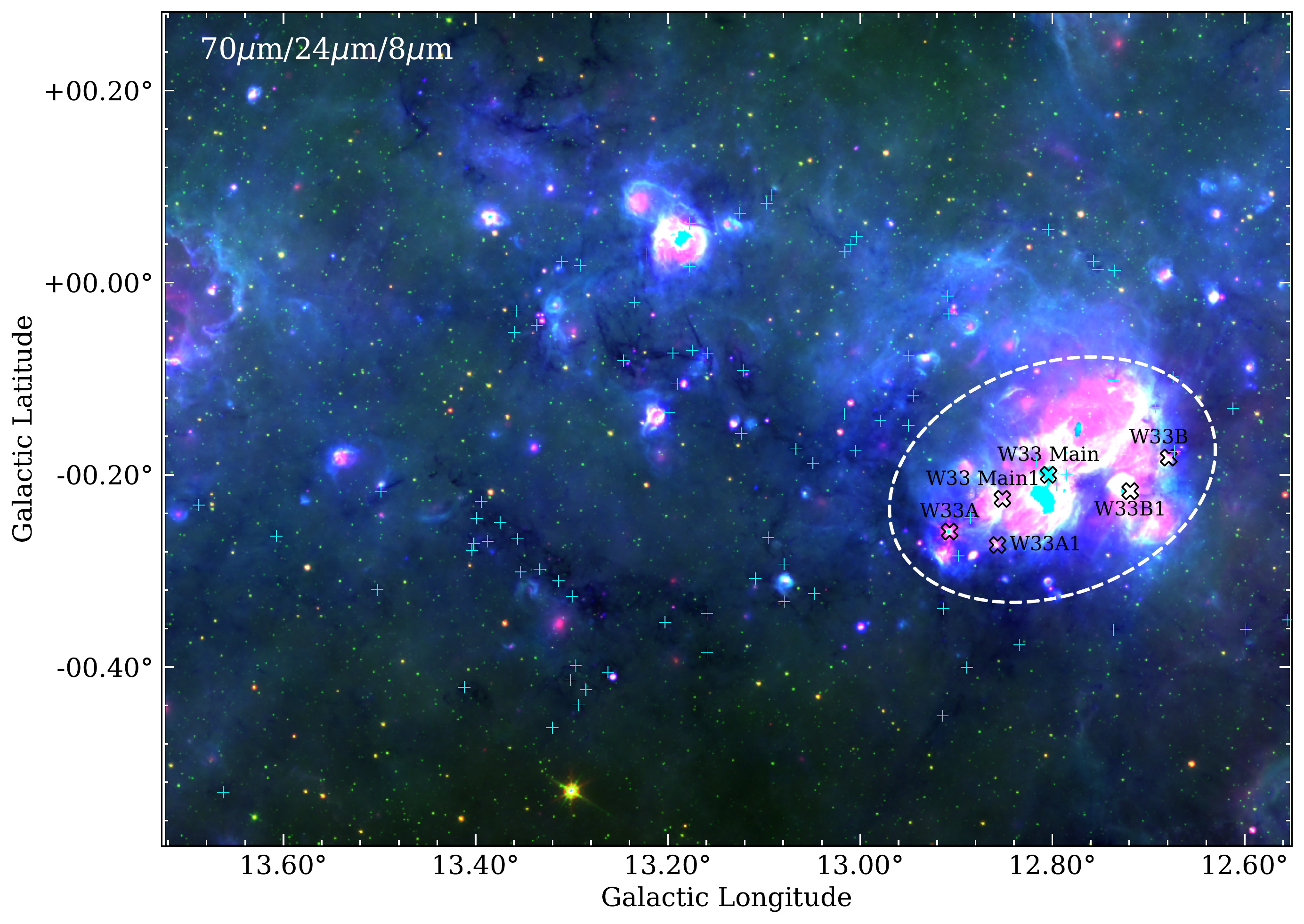}
\caption{Three-olour composite image towards the W33 complex and its surroundings with blue, green, and red corresponding to Hi-GAL 70 \um \citep{Pilbratt10}, GLIMPSE 8.0 \um \citep{Benjamin03}, and MIPSGAL 24 \um \citep{Carey09}, respectively. The `\textit{Xs}' symbols represent the massive clumps W33 Main, W33A, W33B, W33 Main1, W33A1, and W33B1 identified by \citet{Immer14} with the ATLASGAL survey at 870 \um \citep{Schuller09}. The plus symbols mark the locations of IRDCs identified by \citet{Peretto16}, and the white dashed ellipse represents the W33 complex. }
\end{figure*}

\begin{figure*}
  \centering
   \includegraphics[angle=0,scale=0.45]{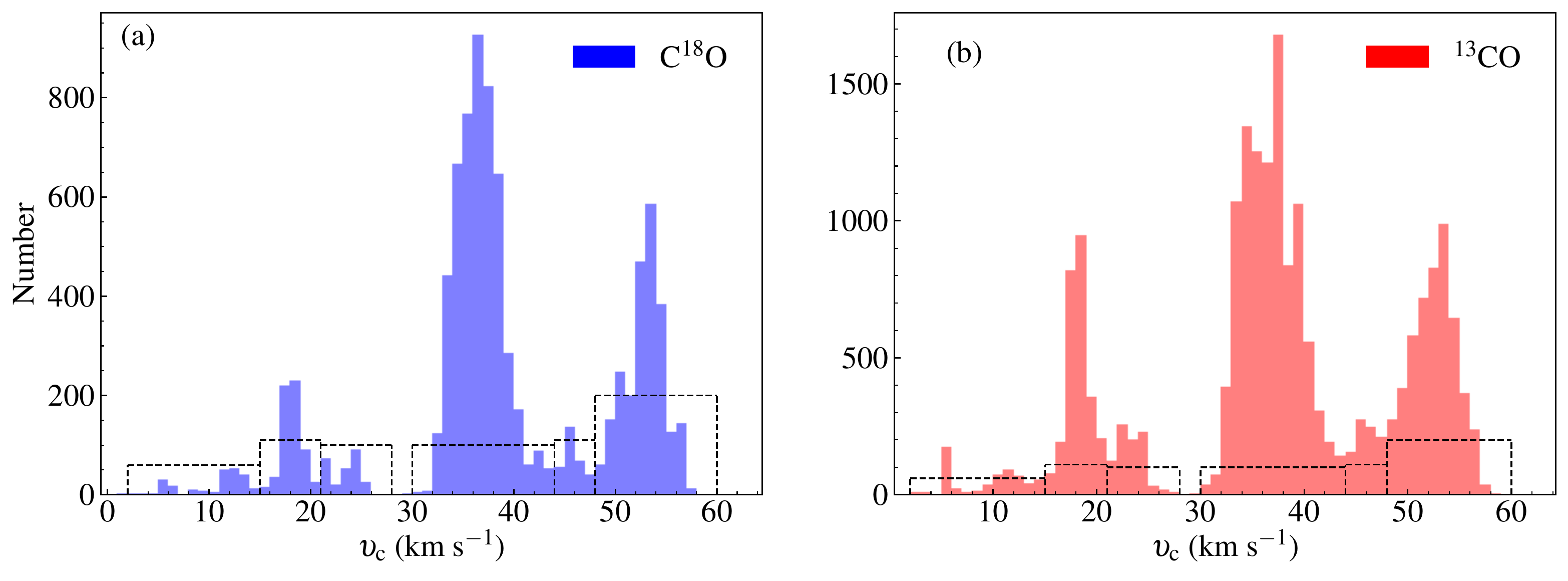}
\caption{Histograms showing the distributions of the velocity centroid of the C$^{18}$O (1-0) and $^{13}$CO (1-0) fits, respectively.}
\end{figure*}

\begin{figure*}
  \centering
   \includegraphics[angle=0,scale=0.6]{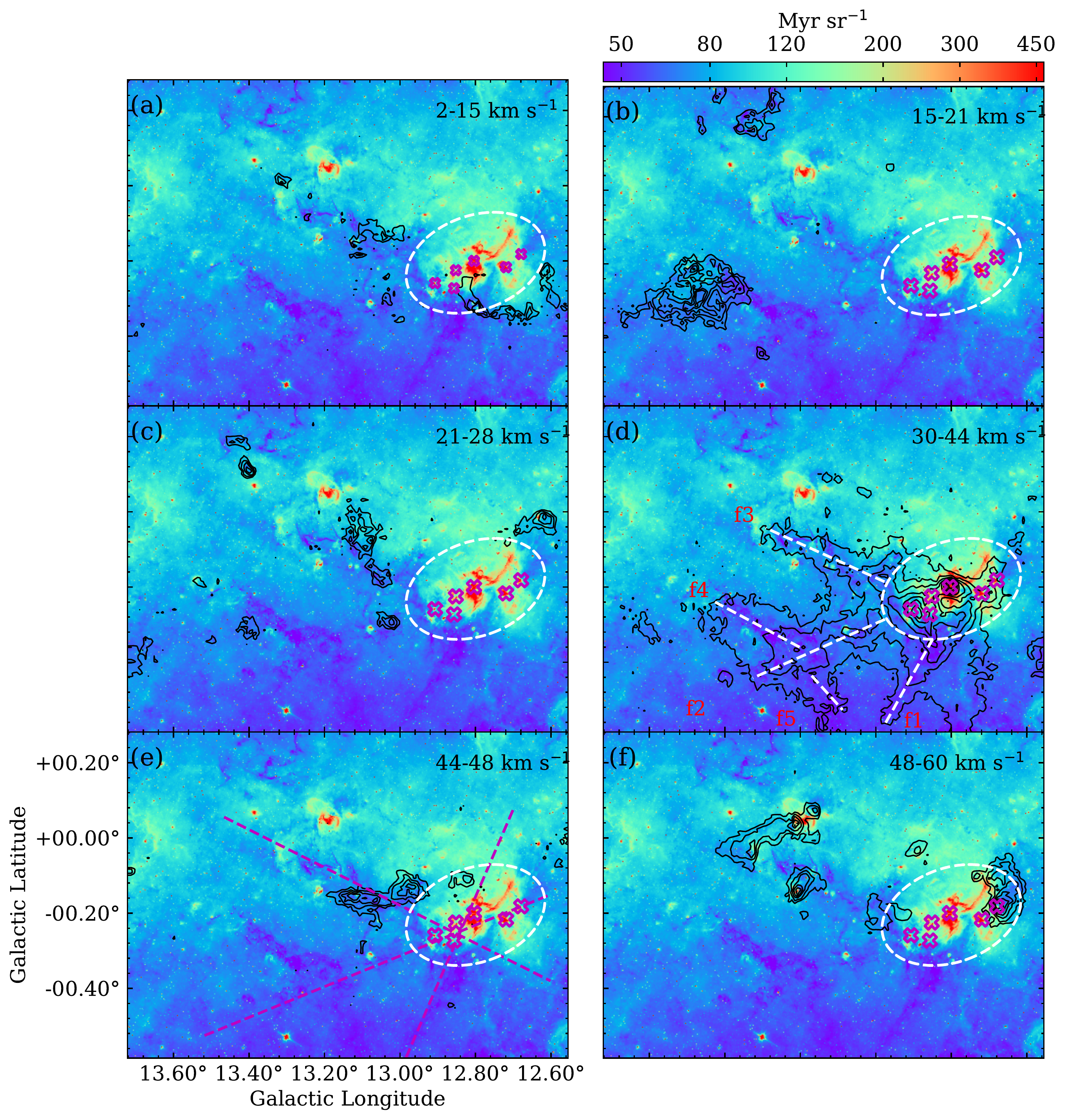}
\caption{(a)-(f) Velocity-integrated C$^{18}$O (1-0) contours in the velocity ranges of 2-15 \kms, 15-21 \kms, 21-28 \kms, 30-44 \kms, 44-48 \kms, and 48-60 \kms respectively, overlaid on the \emph{Spitzer} 8 \um image. The white dashed lines in (d) show the filaments detected in 30-44 \kms, and the magenta dashed lines in (e)  represent the cutting directions of Fig. 10. The colour bar represents the flux at 8 \um in units of Myr sr$^{-1}$.}
\end{figure*}

\begin{figure*}
  \centering
   \includegraphics[angle=0,scale=0.3]{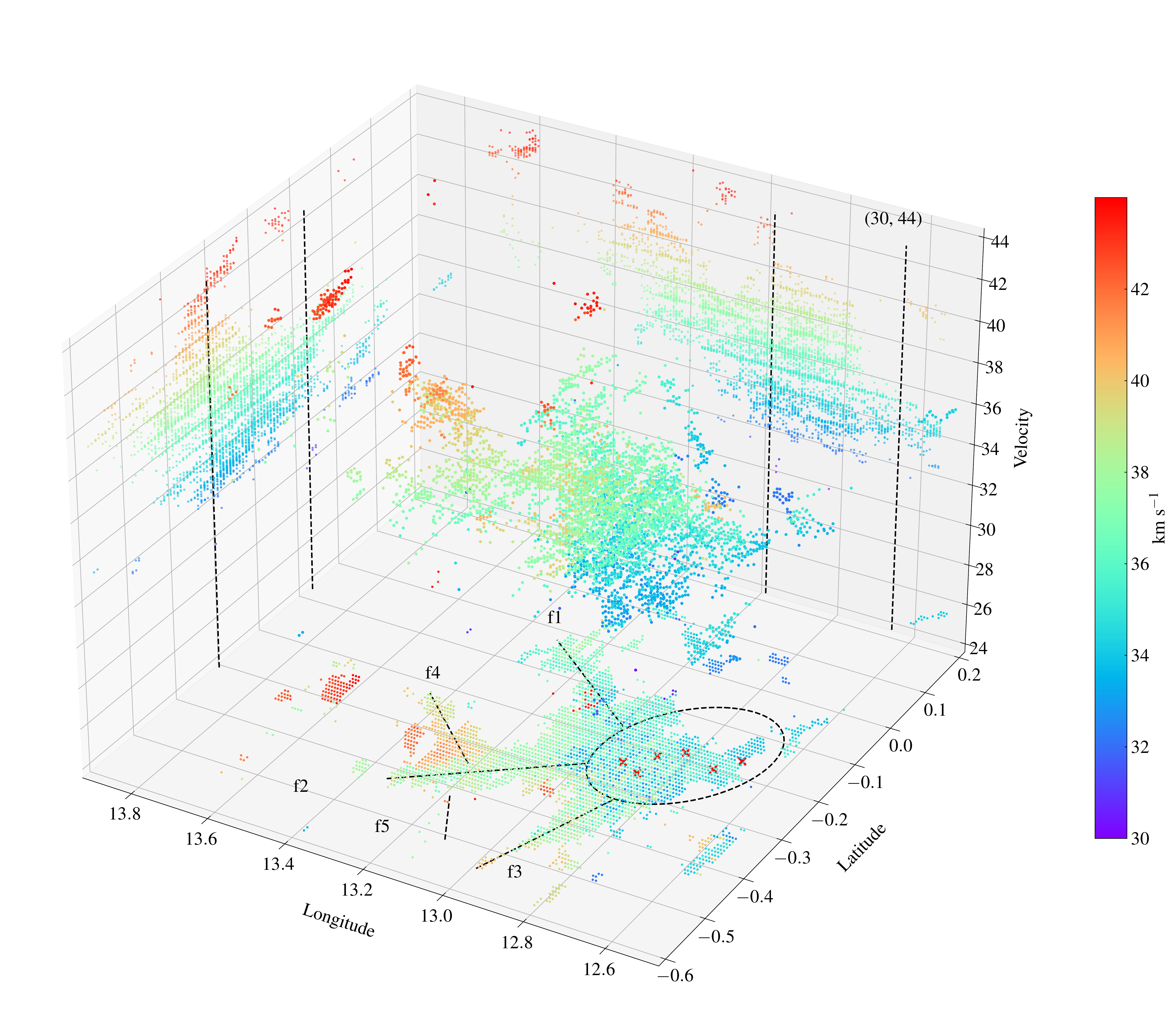}
\caption{The C$^{18}$O (1-0) PPV space in the velocity interval of 30-44 \kms. The projections on the three axes are presented. The colour of each point represents the centroid velocity at that point, corresponding to the colour bar shown to the right of the panel. The symbols `\textit{Xs}' are similar to those in Fig. 1, and the W33 complex is marked by the black dashed lines or the dashed ellipses on the three projections. }
\end{figure*}

\clearpage
\begin{figure*}
  \centering
   \includegraphics[angle=0,scale=0.6]{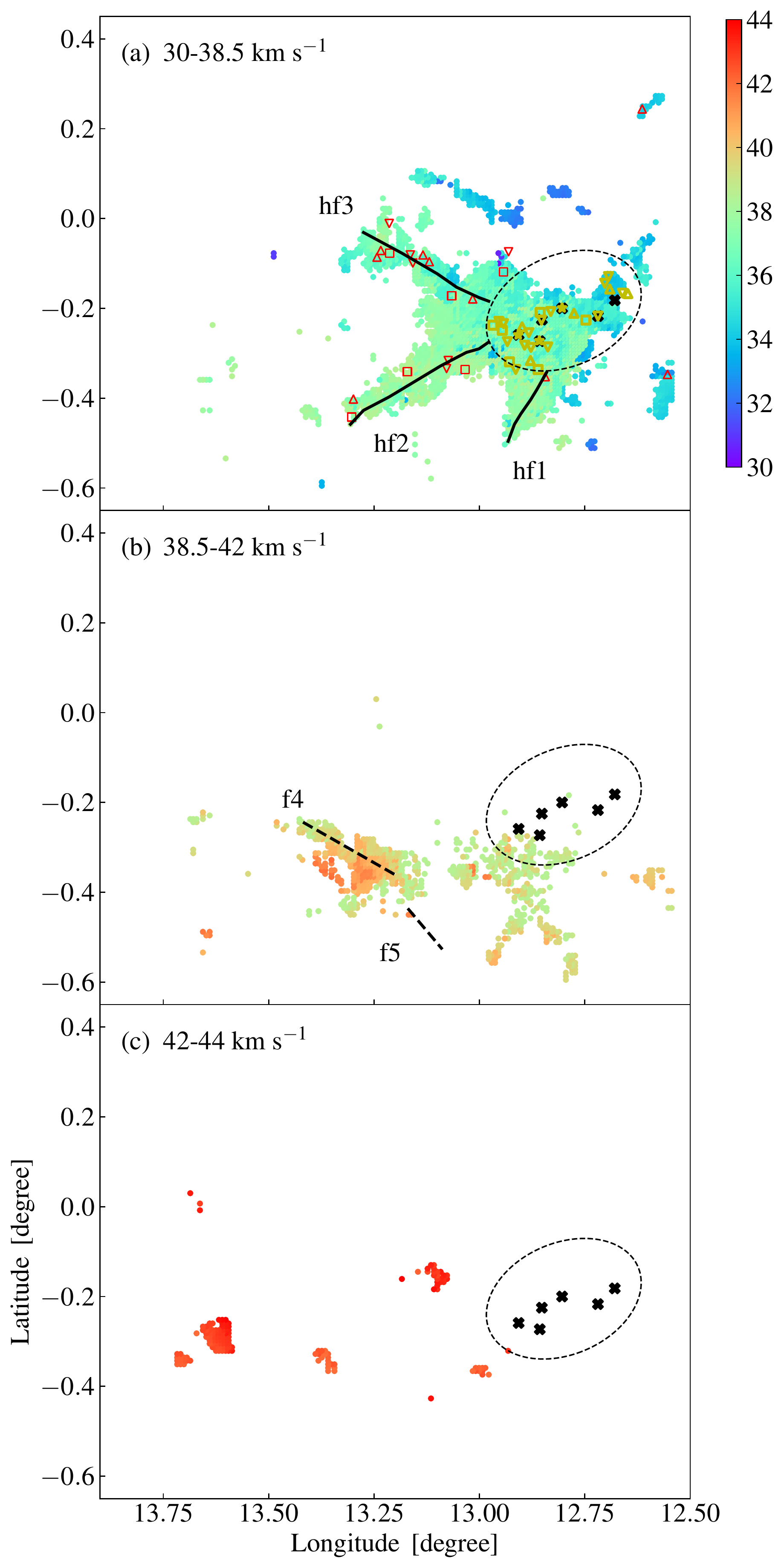}
\caption{Plots showing the velocity distributions in 30-38.5 \kms (a), 38.5-42 \kms (b), and 42-44 \kms (c) respectively. The points are coloured with the velocity values shown in the right colour bar. The asterisks represent the MSF clumps, `$\bigtriangledown$' mark the YSO clumps, `$\bigtriangleup$' flag the protostellar clumps, and the squares are for the quiescent clumps. The black lines denote the spines of the filaments hf1, hf2, and hf3. The clumps are coded with different colours in the W33 complex (yellow) and in the filaments (red), respectively.}
\end{figure*}

\begin{figure*}
  \centering
   \includegraphics[angle=0,scale=0.5]{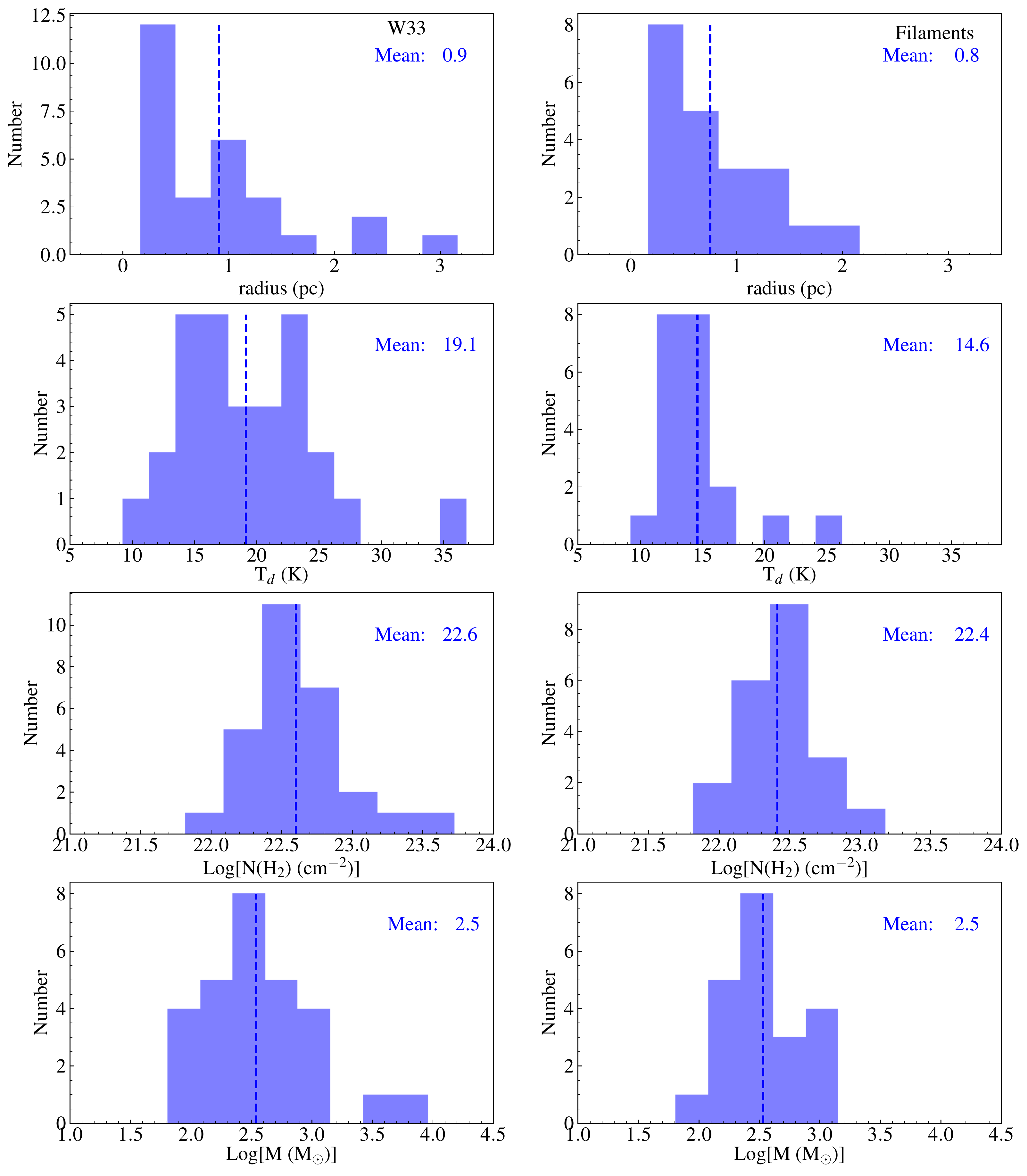}
\caption{Comparisons of the clump physical parameter distributions in the central hub W33 and in the filaments.}
\end{figure*}

\begin{figure*}
\centering
\includegraphics[angle=0,width=0.6\textwidth]{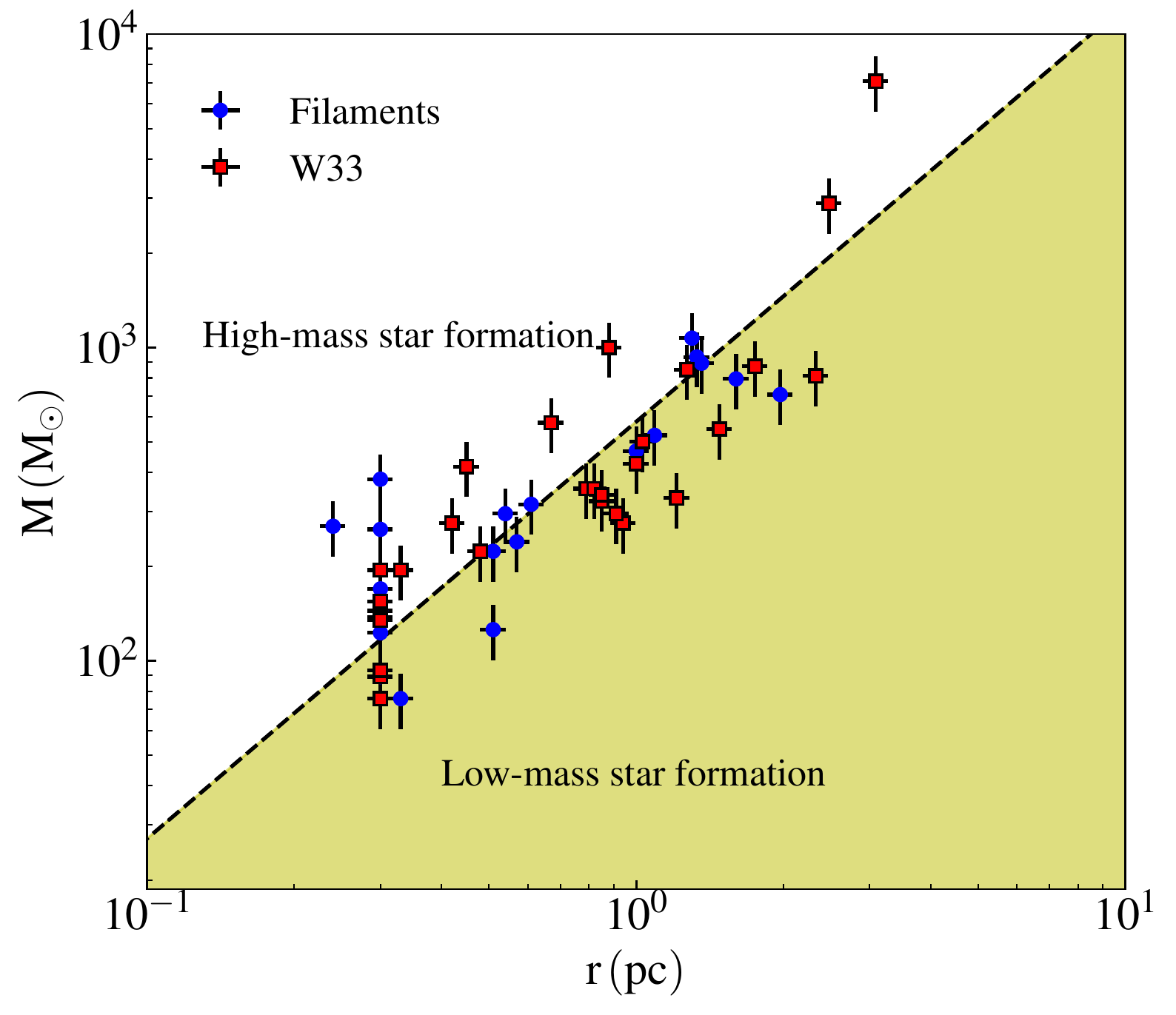}
\caption{Mass--size relationship of the clumps with masses determined. The yellow shaded region represents the parameter space devoid of massive star formation, where $ M/\rm {M_\odot}=580\,(R \,\rm{pc^{-1}})^{1.33}$ \citep{Kauffmann10}. The red squares represent the clumps in the central hub W33 and the blue circlesmark the clumps in the filaments. The errors on the masses and radii of the clumps are $\sim 20\%$ and $\sim 6\%$, respectively, from \citet{Urquhart18}. }
\end{figure*}

\begin{figure*}
\centering
\includegraphics[angle=0,width=0.6\textwidth]{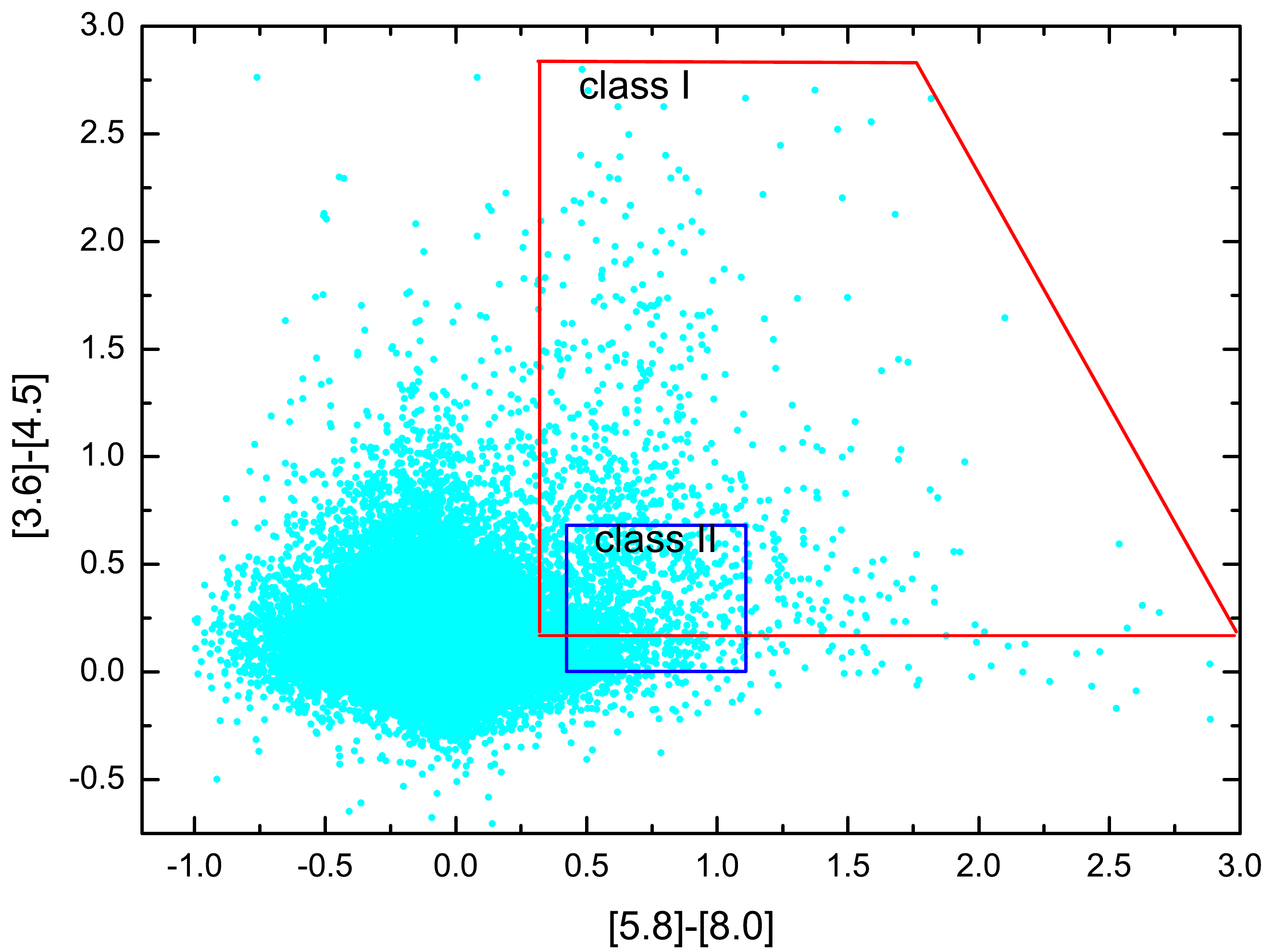}
\caption{IRAC colour-colour plot for the sources in our observational region. The regions indicated the stellar evolutionary stage as defined by \citet{Allen04}. Class I sources represent protostars with circumstellar envelopes and Class II are disc-dominated objects. }
\end{figure*}

\begin{figure*}
\centering
\includegraphics[angle=0,width=0.9\textwidth]{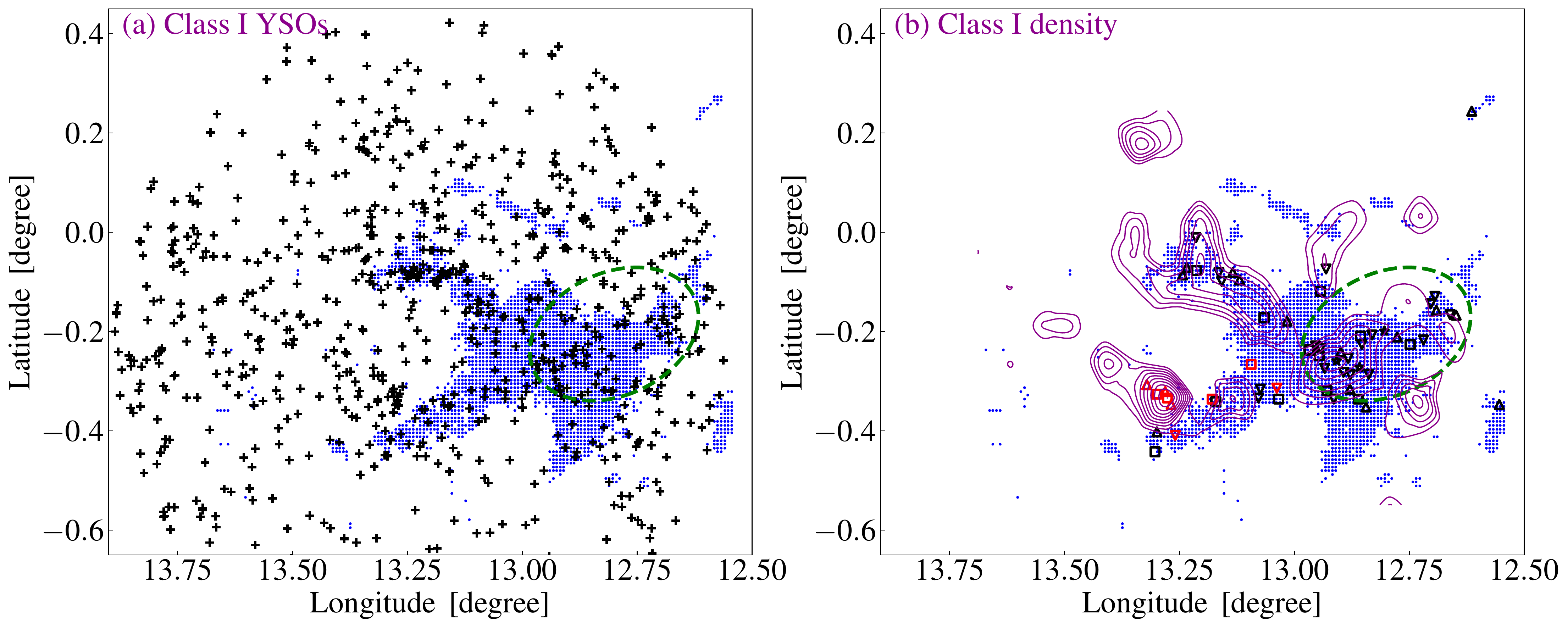}
\caption{(a) Class I YSOs marked by plus symbols overlaid on the C$^{18}$O (1-0) emission. (b) Corresponding overlay of Class I YSOs surface density contours on the C$^{18}$O (1-0) emission. The contour levels start from 0.5 pc$^{-2}$ to 1.7 pc$^{-2}$ (magenta contours) in steps of 0.13 pc$^{-2}$. The different symbols represent the clumps in different stages, similar to those in Fig. 5a. In addition, the clumps marked with the red colour are in the velocity range of 38.5-42 km s$^{-1}$.}
\end{figure*}

\begin{figure*}
   \begin{minipage}[t]{0.33\textwidth}
  \centering
   \includegraphics[angle=0,scale=0.3]{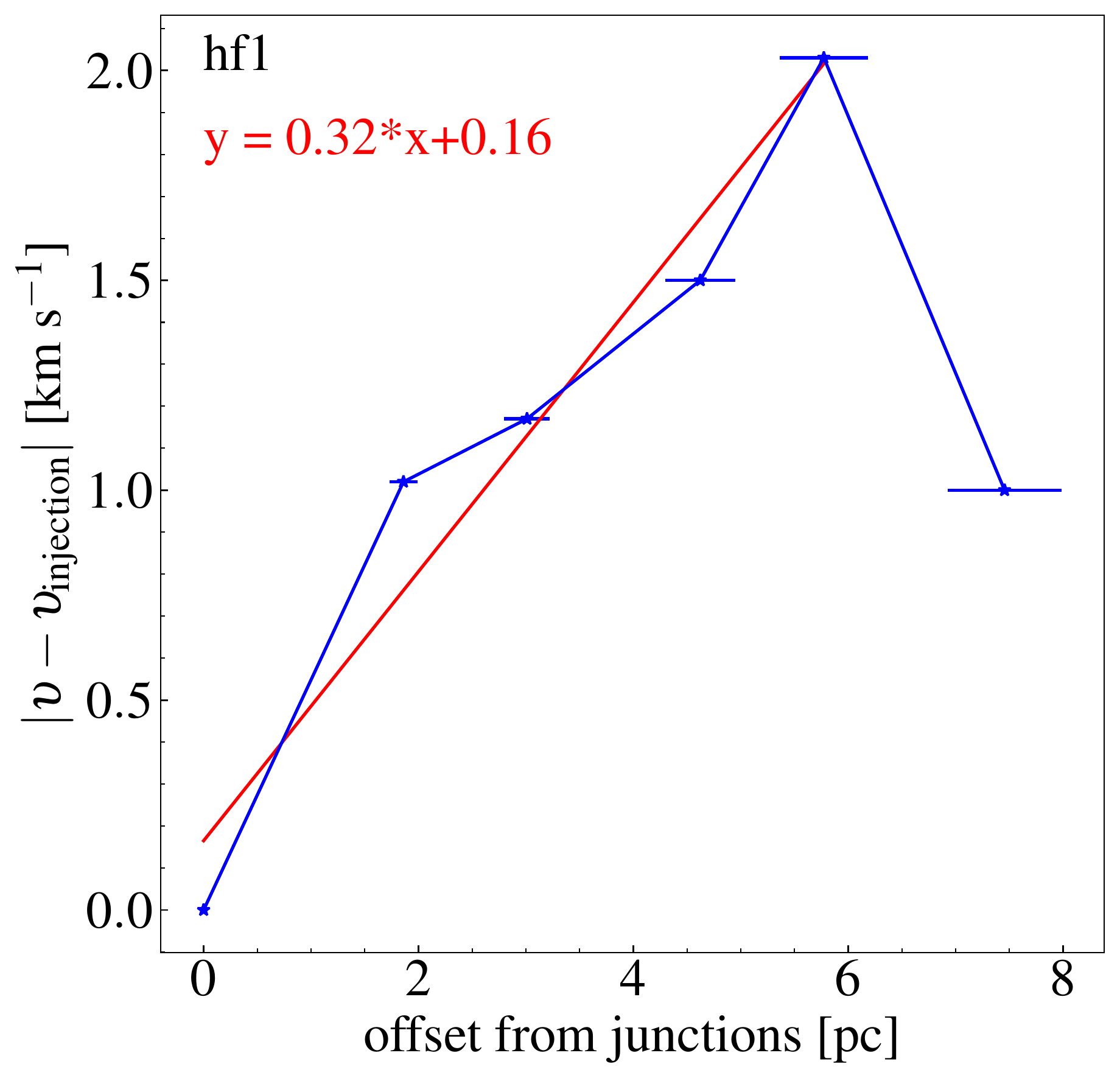}
  \end{minipage}%
  \begin{minipage}[t]{0.33\textwidth}
  \centering
   \includegraphics[angle=0,scale=0.3]{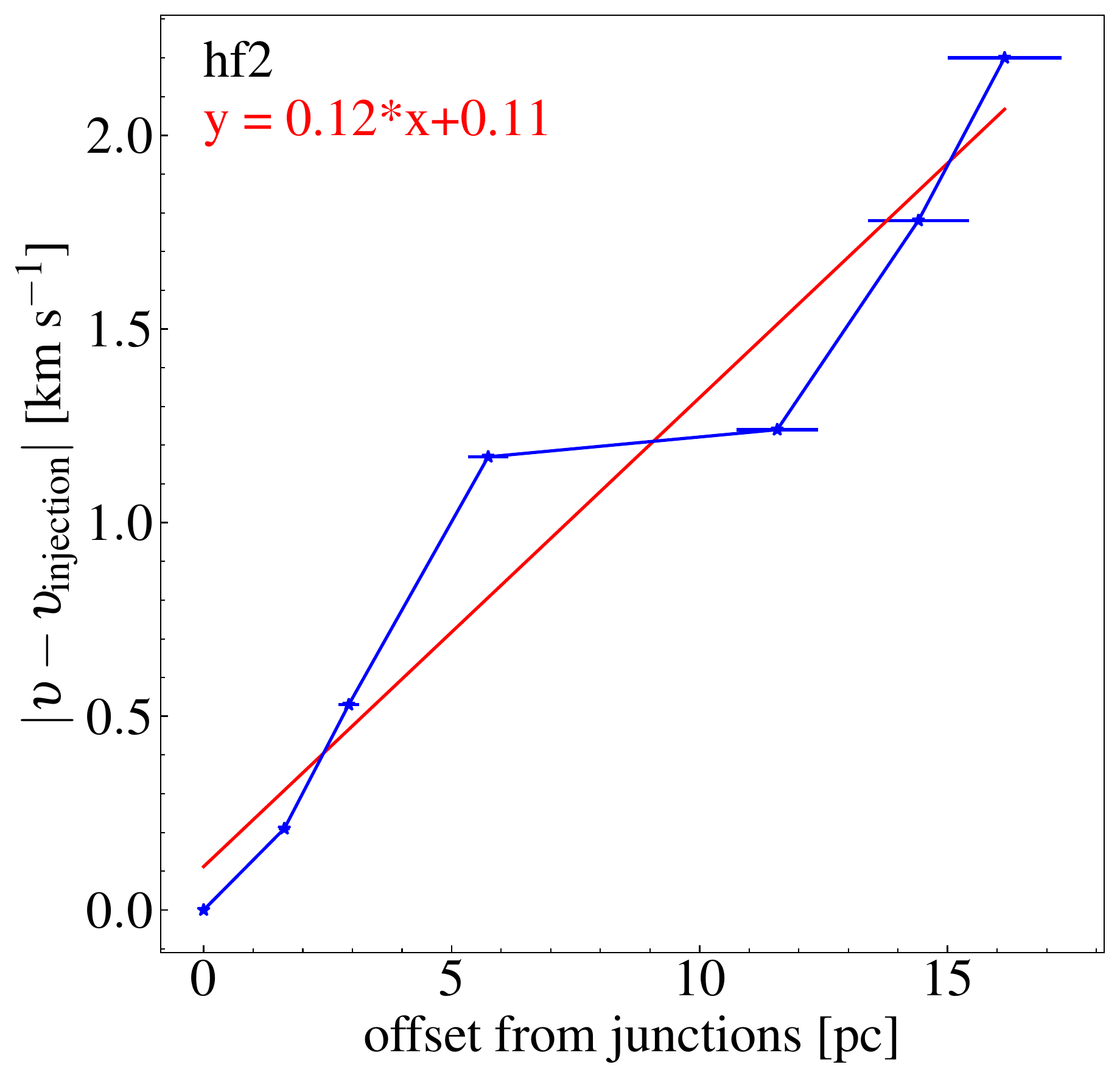}
  \end{minipage}%
  \begin{minipage}[t]{0.33\textwidth}
  \centering
   \includegraphics[angle=0,scale=0.3]{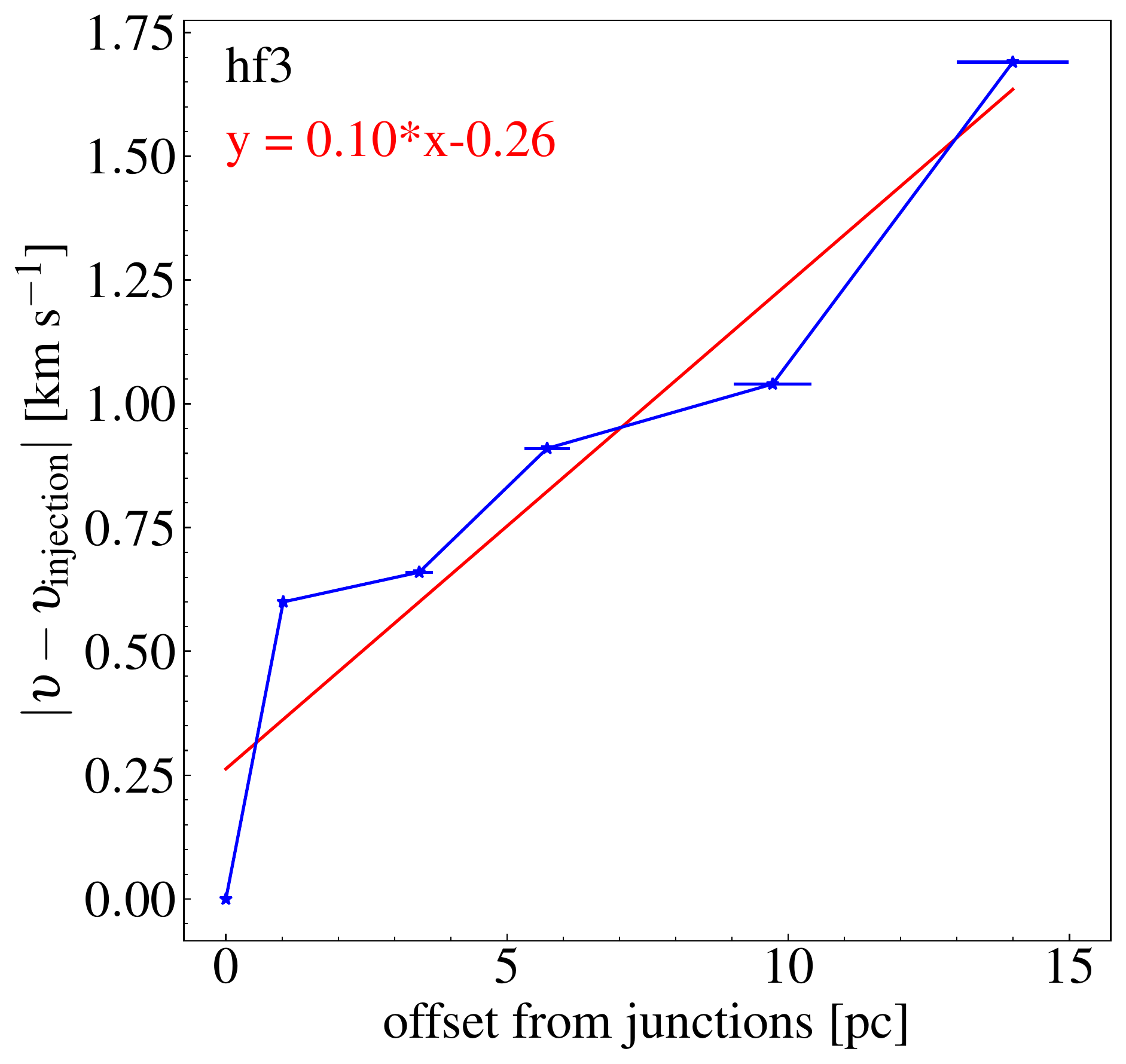}
  \end{minipage}%

\caption{Line-of-sight velocity difference of C$^{18}$O (1-0) as a function of position from junctions.}
\end{figure*}

\begin{figure*}
\centering
\includegraphics[angle=0,width=0.8\textwidth]{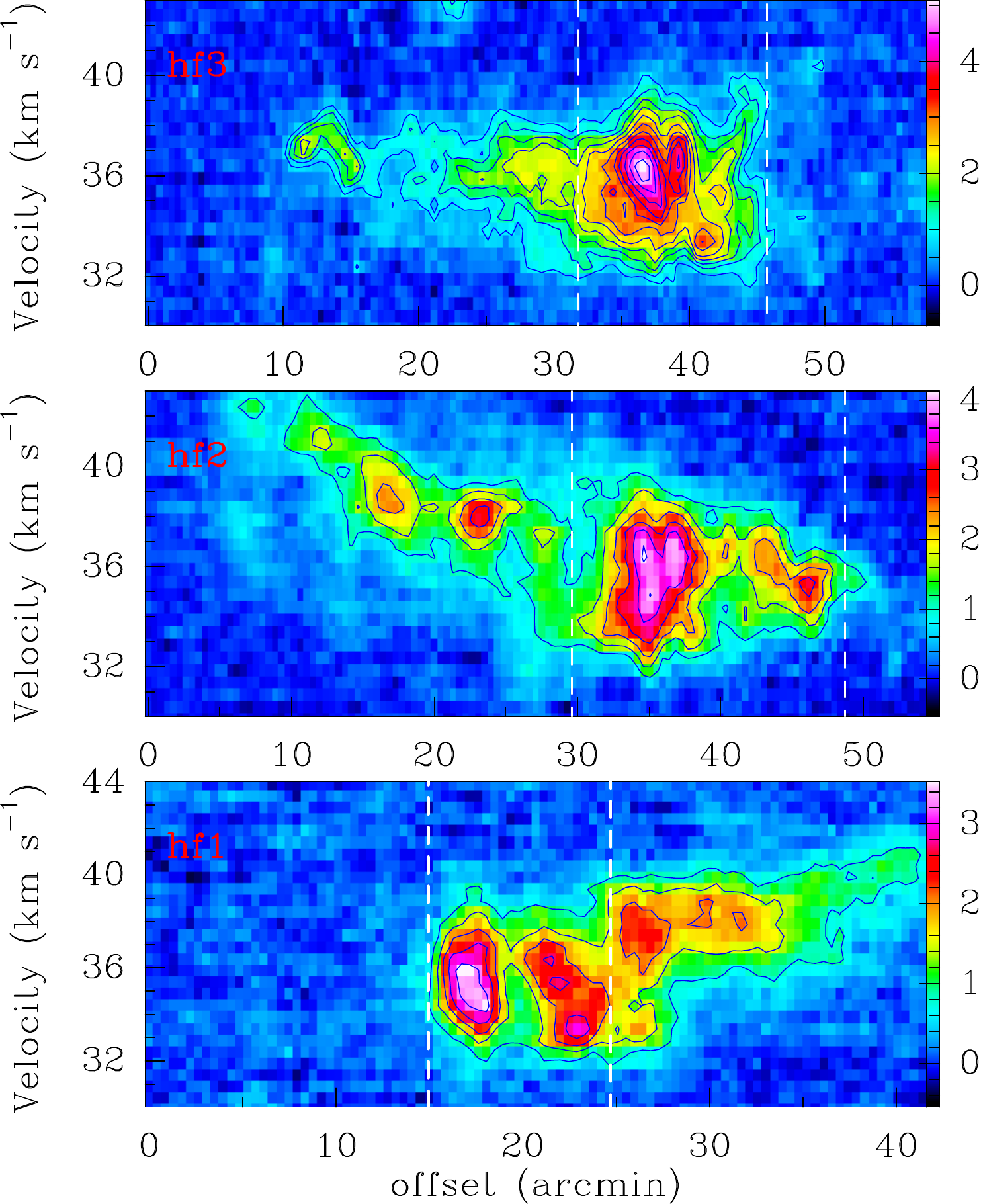}
\caption{Position--velocity diagrams of C$^{18}$O (1-0) along the cuts in Fig. 3e. The contour levels start from 5$\sigma$ to the peak integrated intensity by 3$\sigma$ ($\sigma = 0.2$ K). The vertical white dashed lines mark the location of the W33 complex, and the filaments detected are labelled in each panel.}
\end{figure*}

\begin{appendix}
\section{Calculations to derive the physical parameters of each fit}
In general, the $^{12}$CO (1-0) emission is optically thick. Therefore, we can estimate the excitation temperature \Tex for each fit of $^{13}$CO (1-0) and C$^{18}$O (1-0) respectively via the flowing formula \citep{Garden91, Pineda08}
\begin{equation}
T_{\rm ex}=\frac{5.53}{ln[1+5.53/(T_{\rm mb}(\rm^{12}CO)+0.82)]},
\end{equation}
where $T_{\rm mb}(\rm^{12}CO)$ is the $^{12}$CO (1-0) peak intensity in the velocity range ($\upsilon\rm_c-1.175\sigma_{\rm c}$, $\upsilon\rm_c+1.175\sigma_{\rm c}$) of each fit from the $^{13}$CO (1-0) and C$^{18}$O (1-0) spectra, respectively. Because we check and correct every Gaussian fit from the BST code, the median errors for the peak intensity, the centroid velocity $\upsilon\rm_c$ , and the velocity dispersions $\sigma_{\rm c}$ are about $1\%$. Therefore, the derived median excitation temperatures for the $^{13}$CO (1-0) fits and C$^{18}$O (1-0) fits are about $1.7\%$. 

Therefore, we can derive the non-thermal velocity dispersion $\sigma_{\rm NT}$ and the ratio of $\sigma_{\rm NT}$ to the sound speed $c{\rm_s}$. The non-thermal velocity dispersion $ \sigma_{\rm NT}$ of each fit is calculated using the following equation: $\sigma_{\rm NT}=\sqrt{{\sigma_{\rm c}}^2-\frac{k{\rm_B} T_{\rm k}}{\rm m_{gas}}}$, where $\sigma_{\rm c}$ is the dispersion of each fit, $k_{\rm B}$ is the Boltzmann constant, $T_{\rm k}$ is the kinetic temperature (here $T{\rm_k} \approx T_{\rm ex}$ under the assumption of local thermodynamic equilibrium (LTE) \citep{Morgan10}),  $\rm m_{gas}$ is the mass of the molecule, and the sound speed $c_{\rm s}$ can be obtained as $ c{\rm_s}=\sqrt{k{\rm_B} T_{\rm k}/\mu_{\rm H_2} {\rm m_H}}$, in which $\rm m_H$ is the mass of a hydrogen atom and $\rm \mu_{H_2}=2.8$ is the mean molecular weight per hydrogen molecule \citep{Kauffmann08}. Thereby, the median errors of $\sigma_{\rm NT}$ and $\sigma_{\rm NT}/c{\rm_s}$ are $\sim 1.1\%$ and $\sim 1.4\%$, resulting from the errors of the excitation temperature and the velocity dispersion. 

Furthermore, assuming the same excitation temperature between the $^{12}$CO (1-0) line and the C$^{18}$O (1-0) lines, we can estimate the optical depth $\tau$ of the C$^{18}$O (1-0) lines, the H$_2$ column density, the H$_2$ number density, and the mass of each fit. The optical depth $\tau$ of the C$^{18}$O (1-0) lines can be derived from the equation as follows \citep{Garden91,Pineda08}:
\begin{equation}
\tau ={\rm-ln}[1-\frac{T_{\rm mb}}{T_0(J(T_{\rm ex})-J(T_{\rm bg}))}],
\end{equation}
where $T_{\rm mb}$ is the amplitude (i.e. peak intensity) of each fit, $T_0 = h\nu/k$, $h$ is the Planck constant, $k$ is the Boltzmann constant, and $\nu$ is the transition frequency of the optically thin line. Also, $J(T)$ is defined by $J(T) = 1/(\exp(T_0/T)-1)$. The median uncertainty of the optical depth is about $3\%$, caused by the errors of the excitation temperature and the peak intensity.

We can then determine the column density of the optically thin molecular line under the assumption of LTE via the expression \citep{Sanhueza12}:
\begin{equation}
 N =\frac{3k^2}{16\pi^3{\mu_{\rm_D}}^2hB^2}\frac{T_{\rm ex}+hB/3k}{(J+1)^2}\frac{\exp(E_{\rm u}/kT_{\rm ex})}{\exp(h\nu/kT_{\rm ex})-1} \frac{1}{J(T_{ex})-J(T_{bg})}\frac{\tau}{1-\exp(-\tau)} W,
\end{equation}
where $B$ is the rotational constant of the molecule, $J$ is rotational quantum number of the lower state, $\rm \mu_{_D}$ is the permanent dipole moment of the molecule, $\rm T_{bg}=2.73$ is the background temperature, and $E\rm_u$ is the energy of the upper level. The values of these can be obtained from the Cologne Database for Molecular Spectroscopy \footnote{https://cdms.astro.uni-koeln.de/cdms/portal/} \citep[CDMS;][]{Muller01, Muller05}. Also, W is the integrated intensity, which can be estimated using  $\rm FWHW \times amplitude$. Next, the H$_2$ column density of each component can be calculated by $ N{\rm(H_2)}/N{\rm(C^{18}O)} \approx 6\times10^6$ for dense regions \citep{Frerking82}. The median uncertainty of the H$_2$ column density is estimated to be $2\%$  using the median errors of the excitation temperature, the optical depth, the centroid velocity, the velocity dispersion, and the peak intensity. The H$_2$ number density and the mass have the same uncertainties as the H$_2$ column density with a median value of $2\%$.

Finally, we calculate the H$_2$ number density and mass of each component by assuming the point as a rectangle shape. The equations are as follows:
 \begin{equation}
n{\rm( H_2)} = N{\rm(H_2)}/R\rm_{pixel},
\end{equation}

\begin{equation}
M = \mu_{\rm H_2} {\rm m_H} N{\rm(H_2)}(R\rm_{pixel})^2,
\end{equation}
where $ R\rm_{pixel}$ is the size of a pixel, $\rm \mu_{H_2}=2.8$ is the mean molecular weight per hydrogen molecule, and $\rm m_H$ is the mass of a hydrogen atom.

\begin{figure*}
  \centering
   \includegraphics[angle=0,scale=0.5]{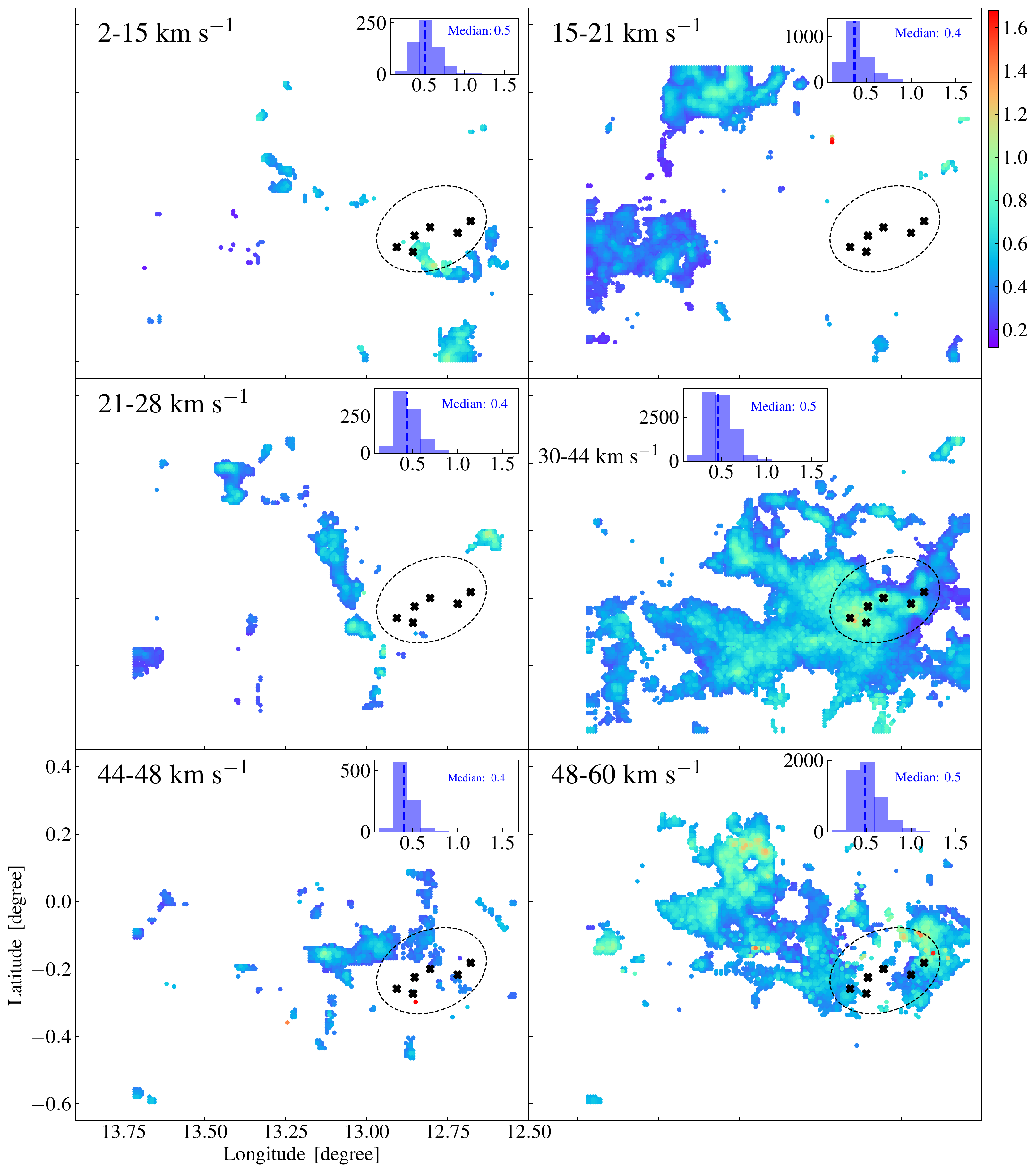}
\caption{Distributions of the optical depth of the $^{13}$CO (1-0) line in different velocity components. The velocity interval of each panel is shown in the left-top corner. The dashed ellipse in each panel marks the W33 complex \citep{Immer14,Kohno18} and the `\textit{Xs}' symbols are the same as those in Fig. 1. }
\end{figure*}

\begin{figure*}
  \centering
   \includegraphics[angle=0,scale=0.5]{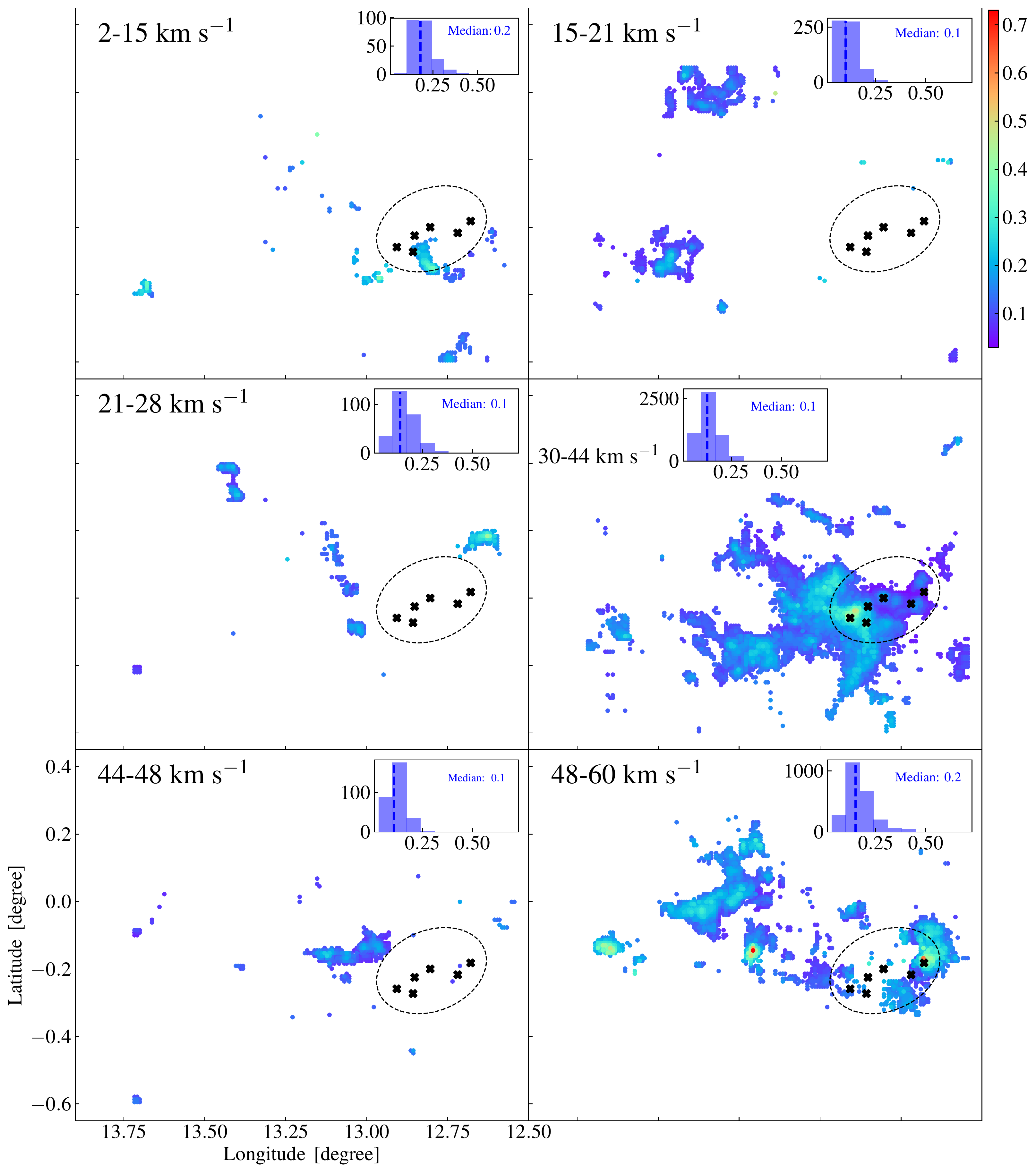}
\caption{Distributions of the optical depth of the C$^{18}$O (1-0) line in different velocity components. The velocity interval of each panel is shown in the left-top corner. The dashed ellipse in each panel marks the W33 complex \citep{Immer14,Kohno18} and the `\textit{Xs}' symbols are the same as those in Fig. 1. }
\end{figure*}

\begin{figure*}
  \centering
   \includegraphics[angle=0,scale=0.5]{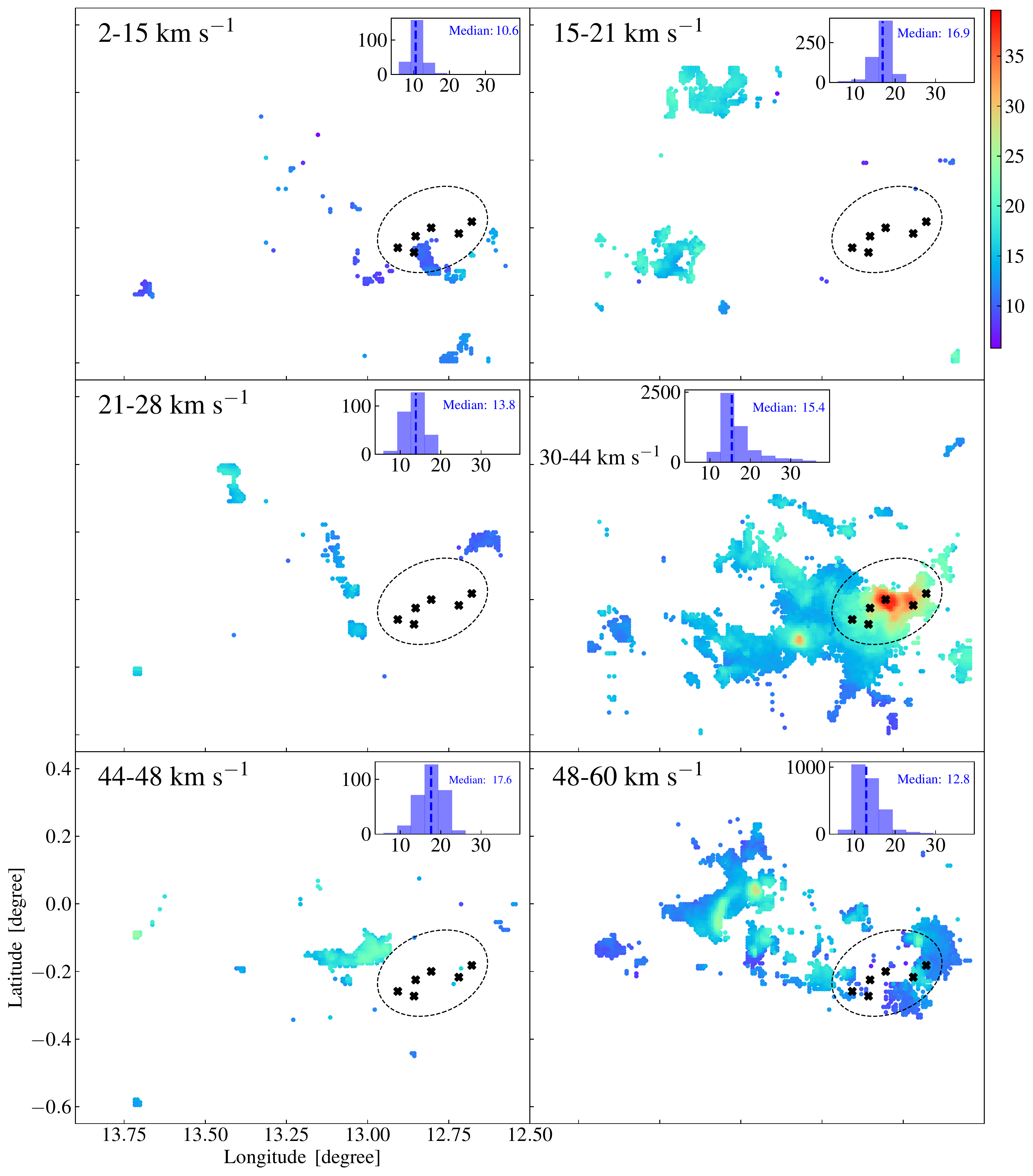}
\caption{Distributions of the excitation temperature in different velocity components. The velocity interval of each panel is shown in the left-top corner. The dashed ellipse in each panel marks the W33 complex \citep{Immer14,Kohno18} and the `\textit{Xs}' symbols are the same as those in Fig. 1. The colour bar is in units of K. }
\end{figure*}

\begin{figure*}
  \centering
   \includegraphics[angle=0,scale=0.5]{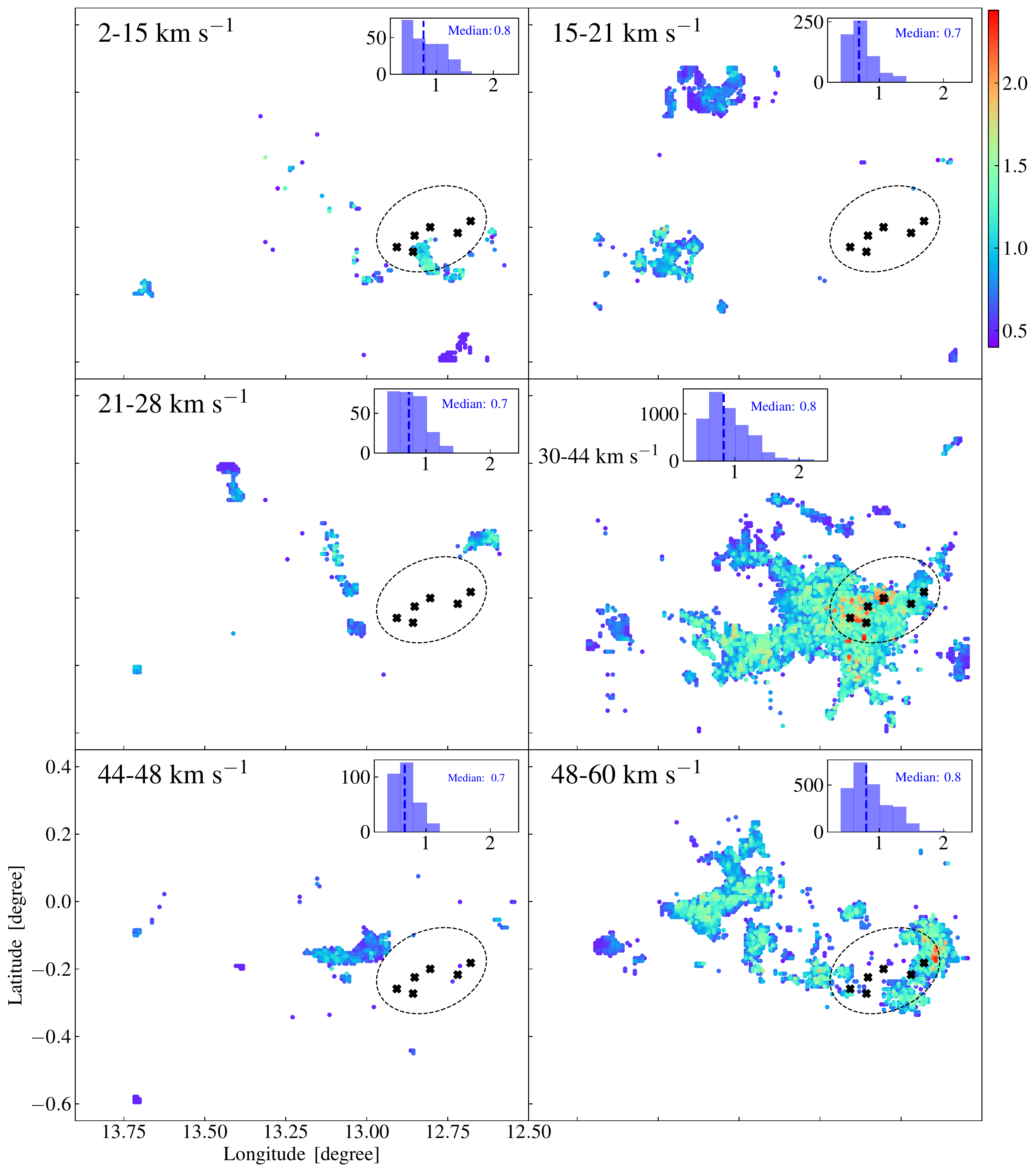}
\caption{Distributions of the C$^{18}$O (1-0) velocity dispersion in different velocity components. The velocity interval of each panel is shown in the left-top corner. The dashed ellipse in each panel marks the W33 complex \citep{Immer14,Kohno18} and the `\textit{Xs}' symbols are the same as those in Fig. 1. The colour bar is in units of \kms. }
\end{figure*}

\begin{figure*}
  \centering
   \includegraphics[angle=0,scale=0.5]{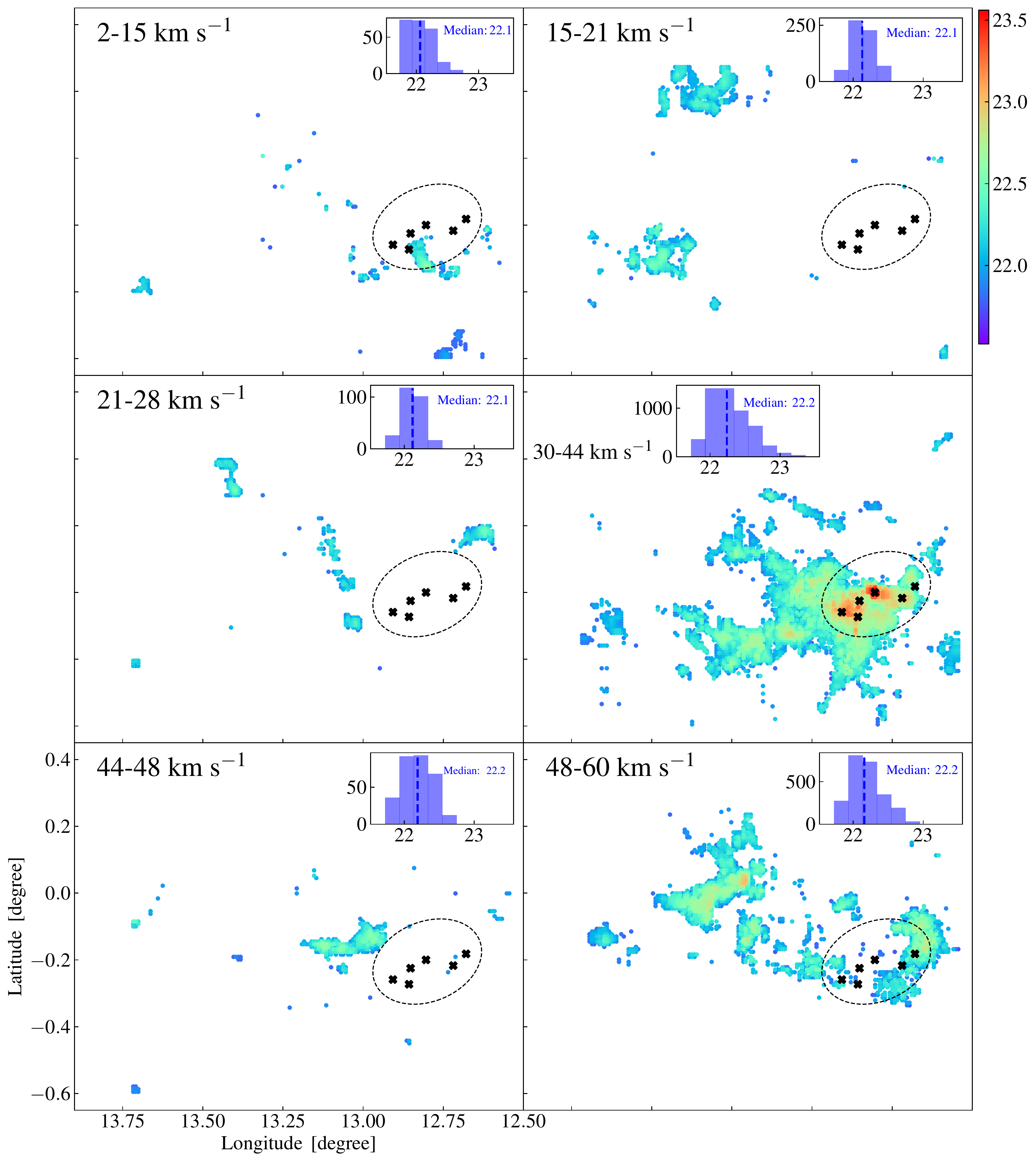}
\caption{Distributions of the H$_2$ column density in different velocity components. The velocity interval of each panel is shown in the left-top corner. The dashed ellipse in each panel marks the W33 complex \citep{Immer14,Kohno18} and the `\textit{Xs}' symbols are the same as those in Fig. 1. The colour bar is logarithmic with units of cm$^{-2}$.}
\end{figure*}

\end{appendix}

\end{document}